\documentstyle[epsfig,12pt]{article}
\newcommand{\be}{\begin{eqnarray}}
\newcommand{\ee}{\end{eqnarray}}

\def\lsim{\mathrel{\rlap{\lower4pt\hbox{\hskip1pt$\sim$}}
    \raise1pt\hbox{$<$}}}               
\def\gsim{\mathrel{\rlap{\lower4pt\hbox{\hskip1pt$\sim$}}
    \raise1pt\hbox{$>$}}}               

\begin{document}

\rightline{\Large{Preprint RM3-TH/01-4}}

\vspace{1cm}

\begin{center}

\LARGE{Leading and higher twists in the proton polarized structure function $g_1^p$ at large Bjorken $x$\footnote{\bf To appear in Physical Review D.}}\\

\vspace{1cm}

\large{S. Simula$^{(a)}$, M. Osipenko$^{(b)}$, G. Ricco$^{(c)}$ and M. Taiuti$^{(c)}$}\\

\vspace{0.5cm}

\normalsize{$^{(a)}$Istituto Nazionale di Fisica Nucleare, Sezione Roma III,\\ Via della Vasca Navale 84, I-00146 Roma, Italy\\$^{(b)}$Physics Department, Moscow State University, 119899 Moscow, Russia\\$^{(c)}$Dipartimento di Fisica, Universit\'a di Genova and INFN, Sezione di Genova,\\ Via Dodecanneso 33, I-16146, Genova, Italy}

\end{center}

\vspace{0.5cm}

\begin{abstract}

\noindent  A phenomenological parameterization of the proton polarized structure function $g_1^p(x, Q^2)$ has been developed for $x \gsim 0.02$ using deep inelastic data up to $\sim 50 ~ (GeV/c)^2$ as well as available experimental results on both photo- and electro-production of proton resonances. According to the new parameterization the generalized Drell-Hearn-Gerasimov sum rule is predicted to have a zero-crossing point at $Q^2 = 0.16 \pm 0.04 ~ (GeV/c)^2$. Then, low-order polarized Nachtmann moments have been estimated and their $Q^2$-behavior has been investigated in terms of leading and higher twists for $Q^2 \gsim 1 ~ (GeV/c)^2$. The leading twist has been treated at $NLO$ in the strong coupling constant and the effects of higher orders of the perturbative series have been estimated using soft-gluon resummation techniques. In case of the first moment higher-twist effects are found to be quite small for $Q^2 \gsim 1 ~ (GeV/c)^2$, and the singlet axial charge has been determined to be $a_0[10 ~ (GeV/c)^2] = 0.16 \pm 0.09$. In case of higher order moments, which are sensitive to the large-$x$ region, higher-twist effects are significantly reduced by the introduction of soft gluon contributions, but they are still relevant at $Q^2 \sim$ few $(GeV/c)^2$ at variance with the case of the unpolarized transverse structure function of the proton. Our finding suggests that spin-dependent correlations among partons may have more impact than spin-independent ones. As a byproduct, it is also shown that the Bloom-Gilman local duality is strongly violated in the region of polarized electroproduction of the $\Delta(1232)$ resonance.

\end{abstract}

\vspace{0.25cm}

PACS numbers: 13.60.Hb, 12.38.-t, 12.38.Cy 

\vspace{0.25cm}

Keywords: \parbox[t]{12cm}{proton structure functions; multiparton correlations; Drell-Hearn-Gerasimov sum rule.}

\newpage

\pagestyle{plain}

\section{Introduction}

\indent The experimental investigation of lepton deep-inelastic scattering ($DIS$) off proton and deuteron targets has provided a wealth of information on parton distributions in the nucleon, leading to a nice confirmation of the leading ($LO$) and next-to-leading order ($NLO$) predictions of the perturbative Quantum Chromodynamics ($QCD$). In the past few years some selected issues in the kinematical regions corresponding to large values of the Bjorken variable $x$ have attracted a lot of theoretical and phenomenological interest; among them one should mention: ~ i) the flavor decomposition of the parton distributions, with particular emphasis on the ratio of $d$- to $u$-quark as $x \to 1$ (see Refs. \cite{Ricco,Bodek}), and ~ ii) the occurrence of power corrections associated to {\em dynamical} higher-twist operators measuring the correlations among partons (see Refs. \cite{Ricco}-\cite{renormalons}). The extraction of the latter is of particular relevance since the comparison with theoretical predictions either based on first-principle calculations (like lattice $QCD$ simulations) or obtained from models of the nucleon structure may represent an important test of the non-perturbative $QCD$ regime.

\indent Various analyses of power-suppressed terms in the world data on the unpolarized nucleon structure functions $F_2^N(x, Q^2)$ and $F_L^N(x, Q^2)$ have been carried out in the past years. They are based either on the choice of a phenomenological ansatz \cite{Ji} or on renormalon-inspired models \cite{Ricco,Bodek,renormalons}, adopting for the leading twist the $LO$ or $NLO$ approximations. The effects of the next-to-next-to-leading order ($NNLO$) corrections have been investigated on $F_2^N(x, Q^2)$ and $R_{L/T}^N(x, Q^2)$ in Ref. \cite{Bodek}, and on the parity-violating structure function $x F_3^N(x, Q^2)$ in Ref. \cite{Kataev}. Very recently \cite{SIM00} the effects of high-order radiative corrections on the extraction of leading and higher twists in the transverse structure function $F_2^N(x, Q^2)$ have been considered adopting soft-gluon resummation techniques \cite{SGR}. It has been shown \cite{SIM00} that the extraction of higher twists at large $x$ is remarkably sensitive to soft gluon effects as well as to the updated $PDG$ value of $\alpha_s(M_Z^2)$ \cite{PDG}. Existing analyses indicate that for $x \gsim 0.7$ dynamical power corrections in $F_2^N(x, Q^2)$, $R_{L/T}^N(x, Q^2)$ and $x F_3^N(x, Q^2)$ are not very large. We want to point out that only in Refs. \cite{Ricco} and \cite{SIM00} the $Q^2$ range of the analyses has been extended down to $Q^2 \sim 1 ~ (GeV/c)^2$ thanks to the use of Nachtmann moments, including in this way the contributions of both the nucleon-resonance regions and the nucleon elastic peak. This is clearly worthwhile not only in order to enhance the sensitivity to power-suppressed terms, but also because of parton-hadron duality arguments \cite{duality}.

\indent The aim of this paper is to extend the twist analysis made in Refs. \cite{Ricco} and \cite{SIM00} to the case of the polarized proton structure function $g_1^p(x, Q^2)$. As in Ref. \cite{Ricco}, we make use of the formalism of the Operator Product Expansion ($OPE$) \cite{OPE} and of the (polarized) Nachtmann moments \cite{Nachtmann}, as defined in Refs. \cite{Wandzura,Uematsu}, in order to disentangle the kinematical target-mass corrections from the dynamical higher-twist effects we are interested in. The evaluation of the Nachtmann moments requires however the knowledge of the polarized structure functions in the whole $x$-range for fixed values of $Q^2$. Therefore, we have developed a new parameterization of $g_1^p(x, Q^2)$, which describes the $DIS$ proton data up to $Q^2 \sim 50 ~ (GeV/c)^2$ and includes a phenomenological Breit-Wigner ansatz able to reproduce the existing electroproduction data in the proton-resonance regions. Our interpolation formula for $g_1^p(x, Q^2)$ has been successfully extended down to the photon point, showing that it nicely reproduces the very recent data \cite{Mainz} on the energy dependence of the asymmetry of the transverse photoproduction cross section as well as the experimental value of the proton Drell-Hearn-Gerasimov ($DHG$) sum rule \cite{DHG}. According to our parameterization of $g_1^p(x, Q^2)$ the generalized $DHG$ sum rule is predicted to have a zero-crossing point at $Q^2 = 0.16 \pm 0.04 ~ (GeV/c)^2$.

\indent  Then, low-order polarized Nachtmann moments have been evaluated and their $Q^2$ behavior has been investigated in terms of leading and higher twists for $Q^2 \gsim 1 ~ (GeV/c)^2$. The leading twist is treated at $NLO$ in the strong coupling constant and it is extracted simultaneously with phenomenological higher-twist terms from our pseudo-data. The effects of higher orders of the perturbative series are estimated using the same soft-gluon resummation technique adopted in case of the analyses of the unpolarized data made in Ref. \cite{SIM00}. The main results of our power correction analysis are as follows. As far as the first moment is concerned, the effects of higher twists are found to be quite small for $Q^2 \gsim 1 ~ (GeV/c)^2$. Moreover,  the singlet axial charge is determined to be $a_0[10 ~ (GeV/c)^2] = 0.16 \pm 0.09$; our extracted value is significantly below the naive quark-model expectation (i.e., compatible with the well-known "proton spin crisis"), but it does not exclude completely a value of the singlet axial charge as large as $\simeq 0.25$, in nice agreement with recent estimates (see, e.g., Ref. \cite{Altarelli}). In case of higher order moments, which are more sensitive to the large-$x$ region, higher-twist effects are significantly reduced by the introduction of soft gluon contributions, but they are still relevant at $Q^2 \sim$ few $(GeV/c)^2$, at variance with the case of the unpolarized transverse structure function of the proton (see Ref. \cite{SIM00}). Our finding suggests that spin-dependent correlations among partons may have more impact than spin-independent ones. As a byproduct, it is also shown that the Bloom-Gilman ($BG$) local duality \cite{BG} is strongly violated in the region of polarized electroproduction of the $\Delta(1232)$ resonance.

\indent The paper is organized as follows. In the next Section the Nachtmann definition of the moments and the $NLO$ approximation for the leading twist  are briefly reminded. In Section 3 a new parameterization of $g_1^p(x, Q^2)$, which describes the $DIS$ regime as well as the photo- and electro-production proton-resonance regions is presented and adopted for the evaluation of the Nachtmann moments. Moreover, the issue of a possible local $BG$ duality among the $DIS$ behaviour of $g_1^p(x, Q^2)$ and suitable local averages in the resonance regions is addressed. Section 4 is devoted to a twist analysis of our pseudo-data at $NLO$, while the inclusion of the effects of high-order radiative corrections is presented in Section 5. Finally, our main conclusions are summarized in Section 6. 

\section{The Nachtmann moments and the Leading Twist at $NLO$}

\indent The complete $Q^2$ evolution of the structure functions can be obtained using the $OPE$ \cite{OPE} of the time-ordered product of the two currents entering the virtual-photon nucleon forward Compton scattering amplitude, viz.
 \be
       T[J(z) ~ J(0)] = \sum_{n, \alpha} ~ f_n^{\alpha}(-z^2) ~ z^{\mu_1} 
       z^{\mu_2} ... z^{\mu_n} ~ O_{\mu_1 \mu_2 ... \mu_n}^{\alpha}
       \label{eq:current}
 \ee
where $O_{\mu_1 \mu_2 ... \mu_n}^{\alpha}$ are symmetric traceless operators of dimension $d_n^{\alpha}$ and twist $\tau_n^{\alpha} \equiv d_n^{\alpha} - n$, with $\alpha$ labeling different operators of spin $n$. In Eq. (\ref{eq:current}) $f_n^{\alpha}(-z^2)$ are coefficient functions, which are calculable in $pQCD$ at short-distance. Since the imaginary part of the forward Compton scattering amplitude is simply the hadronic tensor containing the structure functions measured in $DIS$ experiments, Eq. (\ref{eq:current}) leads to the well-known twist expansion for the Cornwall-Norton ($CN$) moments of the nucleon polarized structure function $g_1^N(x, Q^2)$ (see Refs. \cite{Wandzura,Uematsu}), viz.
 \be
       \overline{M}_n^{(1)}(Q^2) \equiv \int_0^1 dx ~ x^{n - 1} ~ g_1^N(x, 
       Q^2) = \sum_{\tau = 2, even}^{\infty} E_{n \tau}[\mu, \alpha_s(Q^2)] 
       ~ O_{n \tau}(\mu) ~ \left( {\mu^2 \over Q^2} \right)^{{\tau - 2 \over 
       2}}
       \label{eq:CN}
 \ee
where $n = 1, 3, 5, ...$, $\mu$ is the renormalization scale, $O_{n \tau}(\mu)$ are the (reduced) matrix elements of operators with definite spin $n$ and twist $\tau$, containing the information about the non-perturbative structure of the target, and $E_{n \tau}(\mu, Q^2)$ are dimensionless coefficient functions, which can be expressed perturbatively as a power series of the running coupling constant $\alpha_s(Q^2)$.

\indent For massless nucleons only operators with spin $n$ contribute to the $n$-th $CN$ moment (\ref{eq:CN}). When the nucleon mass $M$  is taken into account, operators with different spins can contribute and consequently the higher-twist terms in the expansion of the $CN$ moment $\overline{M}_n^{(1)}(Q^2)$ contain now also target-mass terms, which are of pure kinematical origin and therefore of no physical interest. It has been shown by Nachtmann \cite{Nachtmann} in the unpolarized case and subsequently generalized to the polarized structure functions in Refs. \cite{Wandzura,Uematsu} that even when $M \neq 0$ the moments can be redefined in such a way that only spin-$n$ operators contribute in the $n$-th moment, namely
 \be
       M_n^{(1)}(Q^2) & \equiv & \int_0^1 dx {\xi^{n+1} \over x^2} \left\{ 
       g_1^N(x, Q^2) \left[ {x \over \xi} - {n^2 \over (n + 2)^2} {M^2 x^2 
       \over Q^2} {\xi \over x} \right] \right. \nonumber \\
       & & \left. - g_2^N(x, Q^2) ~ {M^2 x^2 \over Q^2} 
       {4n \over n + 2} \right\}
       \label{eq:M1}
 \ee
where $n = 1, 3, 5, ...$ and
 \be
      \xi = {2x \over 1 + \sqrt{1 + 4 M^2 x^2 / Q^2}}
      \label{eq:csi}
 \ee
is the Nachtmann variable. Note that at variance with the unpolarized case, where the transverse Nachtmann moments involve only the transverse structure function $F_2^N(x, Q^2)$ (see, e.g., Ref. \cite{Ricco}), in the polarized case $M_n^{(1)}(Q^2)$ has to be constructed using the two polarized structure functions $g_1^N(x, Q^2)$ and $g_2^N(x, Q^2)$. Using the {\em experimental} data for the latter ones in the r.h.s. of Eq. (\ref{eq:M1}), the target-mass corrections are exactly canceled out; therefore the twist expansions of the experimental Nachtmann moments $M_n^{(1)}(Q^2)$ contain only {\em dynamical} higher twists, which are the only ones related to the correlations among partons, viz.
 \be
      M_n^{(1)}(Q^2) = \delta\mu_n^{(1)}(Q^2) + \mbox{dynamical higher 
      twists}
    \label{eq:M1_NAT}
 \ee
where $\delta\mu_n^{(1)}(Q^2)$ stands for the $n$-th $CN$ moment of the leading twist contribution. The latter can be written in the following form:
 \be
      \delta\mu_n^{(1)}(Q^2) = \delta\mu_n^{NS}(Q^2) + \delta\mu_n^{S}(Q^2) 
      \label{eq:mu_1}
 \ee
where at $NLO$ \cite{NLO}
 \be
       \delta\mu_n^{NS}(Q^2) & = & {<e^2> \over 2} ~ \delta q_n^{NS}(\mu^2) 
       \left( {\alpha_s(Q^2) \over \alpha_s(\mu^2)} \right)^{\gamma_n^{NS}} 
       \left[ 1 + {\alpha_s(Q^2) \over 2 \pi} \delta C_n^{(q)} \right] 
       \nonumber \\
       & & \left[ 1 + {\alpha_s(Q^2) - \alpha_s(\mu^2) \over 4 \pi} \left( 
       \gamma_n^{1, NS} - {\beta_1 \over \beta_0} \gamma_n^{NS} \right) 
       \right]
      \label{eq:mu_NS}
 \ee
and
 \be
      \delta\mu_n^S(Q^2) & = & {<e^2> \over 2} \left\{ \delta\Sigma_n(Q^2) 
      \left[ 1 + {\alpha_s(Q^2) \over 2 \pi} \delta C_n^{(q)} \right] + 
      2 N_f  {\alpha_s(Q^2) \over 2 \pi} \delta G_n(Q^2) ~ \delta C_n^{(g)} 
      \right\}
      \label{eq:mu_S}
 \ee
In Eq. (\ref{eq:mu_NS}) $\delta q_n^{NS}(\mu^2)$ is the $n$-th moment of the non-singlet ($NS$) polarized quark distribution in the (modified) minimal subtraction ($\overline{MS}$) or Adler-Bardeen ($AB$) scheme evaluated at the renormalization scale $\mu^2$, $\gamma_n^{NS}$ and $\gamma_n^{1, NS}$ are the non-singlet (unpolarized) anomalous dimensions at one- and two-loop levels, respectively, $\delta C_n^{(q)}$ is the $n$-th moment of the quark coefficient function, $\beta_0 = 11 - 2 N_f / 3$, $\beta_1 = 102 - 38 N_f / 3$ and eventually $<e^2> \equiv (1 / N_f) \sum_{i = 1}^{N_f} e_i^2$, with $N_f$ being the number of active flavors. In Eq. (\ref{eq:mu_S}) $\delta\Sigma_n(Q^2)$ and $\delta G_n(Q^2)$  are the $n$-th moments of the singlet-quark and gluon polarized distributions, respectively, and $\delta C_n^{(g)}$ is the $n$-th moment of the gluon coefficient function.

\indent As it is well known, the evolutions of the singlet-quark and gluon distributions are coupled in general; however, as it happens in the unpolarized case (cf., e.g., \cite{Ricco}), the moments of order $n > 1$ are sensitive to the large-$x$ regions, where the evolution of gluons and quarks are approximately decoupled. Therefore, in what follows we will assume the following $NLO$ evolution for $\delta \mu_{n \geq 3}^{(1)}$:
 \be
       \delta\mu_{n \geq 3}^{(1)}(Q^2) & = & \delta A_n(\mu^2) 
       \left( {\alpha_s(Q^2) \over \alpha_s(\mu^2)} \right)^{\gamma_n^{NS}} 
       \left[ 1 + {\alpha_s(Q^2) \over 2 \pi} \delta C_n^{(q)} \right] 
       \nonumber \\
       & & \left[ 1 + {\alpha_s(Q^2) - \alpha_s(\mu^2) \over 4 \pi} \left( 
       \gamma_n^{1, NS} - {\beta_1 \over \beta_0} \gamma_n^{NS} \right) 
       \right]
      \label{eq:mu_large_n}
 \ee
where $\delta A_n(\mu^2) \equiv \delta\mu_n^{(1)}(\mu^2) / [1 + \alpha_s(\mu^2) \delta C_n^{(q)} / 2 \pi]$.

\indent In case of the first moment ($n = 1$), adopting the $AB$ scheme as defined in Ref. \cite{BFR96} and introducing the simplified notation $\Delta\mu^{(1)} \equiv \delta\mu_{n = 1}^{(1)}$, one gets
 \be
       \Delta\mu^{(1)}(Q^2) = {<e^2> \over 2} \left[ \Delta q^{NS} + 
       a_0(Q^2) \right] \left[ 1 - {\alpha_s(Q^2) \over \pi} \right]
       \label{eq:mu_first}
 \ee
where $a_0(Q^2)$ is the (scale dependent) singlet axial charge, given by
 \be
       a_0(Q^2) = \Delta \Sigma - N_f {\alpha_s(Q^2) \over 2 \pi} \Delta 
       G(Q^2)
       \label{eq:a0}
 \ee
In Eqs. (\ref{eq:mu_first}-\ref{eq:a0}) both the non-singlet $\Delta q^{NS}$ and the quark singlet $\Delta \Sigma$ are conserved quantities, while the polarized gluon moment $\Delta G$ is scale dependent (at $NLO$ one has $\Delta G(Q^2) \sim 1 / \alpha_s(Q^2)$, so that the axial singlet charge $a_0$ becomes only slightly scale dependent). Below the charm threshold (i.e., $N_f = 3$) one has
 \be
       \Delta q^{NS} = {3 \over 4} g_A + {1 \over 4} a_8
       \label{eq:Delta_NS}
 \ee
where $g_A = 1.2670 \pm 0.0035$ \cite{PDG} is the nucleon axial coupling constant and $a_8 = 0.579 \pm 0.025$ \cite{CR93} is the octect axial charge (obtained from nucleon and hyperon beta decays under the assumption of $SU(3)$-flavor symmetry). Thus, the expected value for $\Delta q^{NS}$ below the charm threshold is $\Delta q^{NS} = 1.095 \pm 0.007$.

\section{Construction of the polarized Nachtmann moments}

\indent For the evaluation of the Nachtmann moments $M_n^{(1)}(Q^2)$ (Eq. (\ref{eq:M1})) systematic measurements of the structure functions $g_1^N(x, Q^2)$ and $g_2^N(x, Q^2)$ are in principle required in the whole $x$-range at fixed values of $Q^2$. The kinematical coverage of existing data on $g_1^p(x, Q^2)$ is shown in Fig. 1. It can be seen that the available data cover the kinematical range $0.3 \lsim Q^2 ~ (GeV/c)^2 \lsim 60$ with values of  the produced invariant mass $W^2 = M^2 + Q^2 (1 - x) / x$ up to $\sim 300 ~ GeV^2$. While data are scarce below $x \sim 0.02$, the coverage at intermediate and high values of $x$, but still in the $DIS$ regions (i.e. for $W \gsim 2 ~ GeV$), appears to be sufficient for developing an interpolation formula. On the contrary, in the nucleon resonance regions ($W \lsim 2 ~ GeV$) only data for $Q^2 \sim 0.5$ and $\sim 1.2 ~ (GeV/c)^2$ are presently available from the $E143$ experiment \cite{E143}.

\begin{figure}[htb]

\centerline{\epsfysize=10cm \epsfig{file=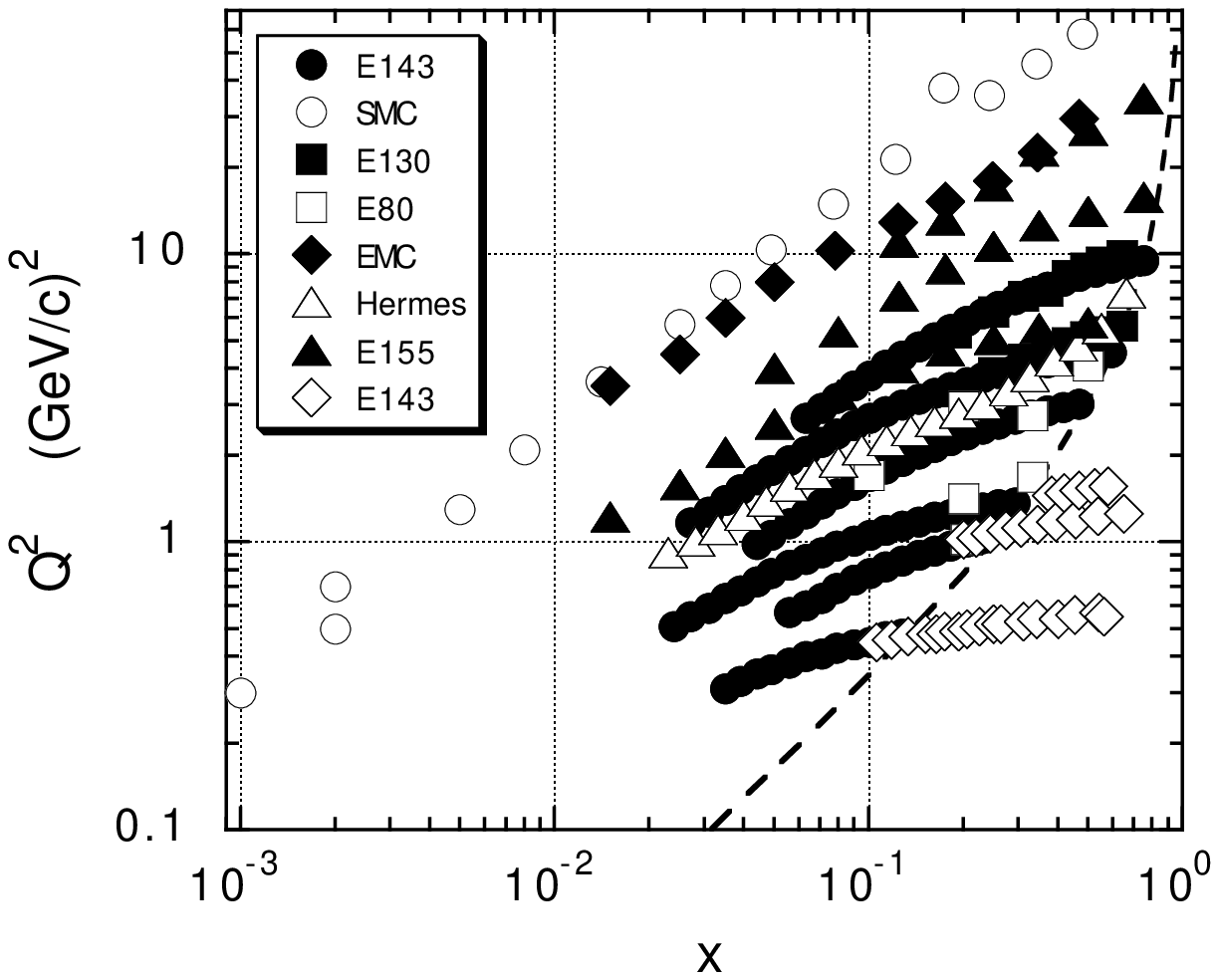}}

\small{ {\bf Figure 1}. Kinematical coverage of existing data on $g_1^p(x, Q^2)$. The dashed line corresponds to an invariant produced mass equal to $W = 2 ~ GeV$. Full dots, open dots, full squares, open squares, full diamonds, open and full triangles  correspond to the experiments of Refs. \cite{E143,SMC,E130,E80,EMC,HERMES,E155}, respectively. Open diamonds are the existing data in the proton-resonance regions from Ref. \cite{E143}.}

\end{figure}

\indent The polarized proton structure functions $g_1(x, Q^2)$ and $g_2(x, Q^2)$ (we omit the suffix $p$ for simplicity) are related to the measured helicity-dependent virtual photon-nucleon cross sections by
 \be
       \label{eq:g1}
       g_1(x, Q^2) & = & {M K \over 4 \pi^2 \alpha_{em}} {1 \over 1 + 4 M^2 
       x^2 / Q^2} \left[ {\sigma_{1/2} - \sigma_{3/2} \over 2} + {2 M x 
       \over Q} \sigma_{LT} \right]  \\
       \label{eq:g2}
       g_2(x, Q^2) & = & {M K \over 4 \pi^2 \alpha_{em}} {1 \over 1 + 4 M^2 
       x^2 / Q^2} \left[ - {\sigma_{1/2} - \sigma_{3/2} \over 2} + {Q \over 
       2 M x} \sigma_{LT} \right] 
 \ee
where $\sigma_{1/2}$ and $\sigma_{3/2}$ are the transverse absorption cross sections for total helicity $1/2$ and $3/2$, respectively, and $\sigma_{LT}$ is the longitudinal-transverse ($LT$) interference cross section. In Eqs. (\ref{eq:g1}-\ref{eq:g2}) $K$ is the incoming flux factor, which in the Hand convention is explicitly given by $K = \nu - Q^2 / 2M = (W^2 - M^2) / 2M$.

\indent Let us write the polarized structure functions as the sum of three contributions, namely 
  \be
       g_i(x, Q^2) = g_i^{(el.)}(x, Q^2) + g_i^{(res.)}(x, Q^2) +  
       g_i^{(non-res.)}(x, Q^2) 
      \label{eq:gi}
 \ee
where $g_i^{(el.)}(x, Q^2)$, $g_i^{(res.)}(x, Q^2)$ and $g_i^{(non-res.)}(x, Q^2)$ are the elastic, resonant and non-resonant contributions to $g_i(x, Q^2)$, respectively, and $i = 1, 2$. In Eq. (\ref{eq:gi}) possible interference terms between the resonant and non-resonant contributions are neglected, since they are well beyond the scope of our phenomenological fit.
The Nachtmann moments (\ref{eq:M1}) can then be written as the sum of three corresponding contributions, viz.
 \be
       M_n^{(1)}(Q^2) = \left[ M_n^{(1)}(Q^2) \right]^{(el.)} + \left[ 
       M_n^{(1)}(Q^2) \right]^{(res.)} + \left[ M_n^{(1)}(Q^2) 
       \right]^{(non-res.)}
      \label{eq:M1_sum}
 \ee

\subsection{Elastic term}

\indent The contribution of the elastic process to the polarized structure functions can be expressed in terms of the electric and magnetic Sachs form factors as
 \be
      \label{eq:g1_el}
      g_1^{(el.)}(x, Q^2) & = & \delta(x - 1) {1 \over 2} G_M(Q^2) {G_E(Q^2) 
      + \tau G_M(Q^2) \over 1 + \tau} \\
      \label{eq:g2_el}
      g_2^{(el.)}(x, Q^2) & = & \delta(x - 1) {\tau \over 2} G_M(Q^2) 
      {G_E(Q^2) - G_M(Q^2) \over 1 + \tau}
 \ee
where $\tau \equiv Q^2 / 4 M^2$. Therefore, one gets
 \be
       \left[ M_n^{(1)}(Q^2) \right]^{(el.)} & = & {\xi_{el}^n \over 2} 
       G_M(Q^2) \left\{ {G_E(Q^2) + \tau G_M(Q^2) \over 1 + \tau} \left[ 1 - 
       {n^2 \over (n + 2)^2} {M^2 \over Q^2} \xi_{el}^2 \right] \right.
       \nonumber \\
       & + & \left. {G_M(Q^2) - G_E(Q^2) \over 1 + \tau} {n \over n + 2} 
       \xi_{el} \right\}
      \label{eq:M1_el}
 \ee
where $\xi_{el} \equiv \xi(x = 1) = 2 / [1 + \sqrt{ 1 + 1 / \tau}]$. For the explicit evaluation of Eq. (\ref{eq:M1_el}) we will make use of the parameterization of Ref. \cite{Drechsel} for the proton elastic form factors, assuming a $5 \%$ uncertainty.

\subsection{Non-resonant contribution}

\indent The non-resonant terms $g_i^{(non-res.)}(x, Q^2)$ can be written in the following form
\be
       g_1^{(non-res.)}(x, Q^2) & = & g^{\Delta \sigma}(x, Q^2) + {4 M^2 x^2 
       \over Q^2} g^{LT}(x, Q^2)  \nonumber \\
       g_2^{(non-res.)}(x, Q^2) & = & - g^{\Delta \sigma}(x, Q^2) + 
       g^{LT}(x, Q^2)
       \label{eq:g12_NR}
 \ee
where $g^{\Delta \sigma}(x, Q^2)$ is the contribution arising from the transverse asymmetry $A_1(x, Q^2)$ [proportional to $(\sigma_{1/2} - \sigma_{3/2})$], while $g^{LT}(x, Q^2)$ is the $LT$ contribution coming from the asymmetry $A_2(x, Q^2)$ [proportional to $\sigma_{LT}$], viz.
 \be
       g^{\Delta \sigma}(x, Q^2) & = & {M K \over 4 \pi^2 \alpha_{em}} {1 
       \over 1 + 4 M^2 x^2 / Q^2} {\sigma_{1/2}^{(non-res.)} - 
       \sigma_{3/2}^{(non-res.)} \over 2} \nonumber \\
       g^{LT}(x, Q^2) & = & {M K \over 4 \pi^2 \alpha_{em}} {1 \over 1 + 4 
       M^2 x^2 / Q^2} {Q \over 2 M x} \sigma_{LT}^{(non-res.)}
       \label{eq:g12_delta_LT}
 \ee

\indent Quite recently \cite{Bianchi}, a description of the transverse cross section difference $(\sigma_{1/2} - \sigma_{3/2})$ in $DIS$ kinematics for both proton and neutron has been obtained in terms of a parameterization inspired by the work of Ref. \cite{ALLM}. There, a simple Regge-type approach, based on one pomeron and one reggeon exchanges, has been shown to be phenomenologically successful in describing both the unpolarized photoproduction cross section and the unpolarized $DIS$ data off the proton. The main result of Ref. \cite{ALLM} is that it is possible to parameterize smoothly the transition from the Regge behaviour, expected to be dominant at low values of $Q^2$, to the partonic description valid at high values of $Q^2$.   Thus, one can try to use the same kind of parameterization to describe $g^{\Delta \sigma}(x, Q^2)$. That was the aim of Ref. \cite{Bianchi}, but unfortunately the explicit form of the parameterization adopted in \cite{Bianchi} fails just in the abovementioned smooth transition, because it does not possess a well defined Bjorken limit.  Therefore, we have developed a new interpolation formula, properly inspired to the work of Ref. \cite{ALLM}, viz.
 \be
       g^{\Delta \sigma}(x, Q^2) = {W^2 - M^2 \over 2W^2} \sum_{j = 1}^N 
       a_j(t) \left[ 1 + {W^2 \over Q^2 + Q_R^2} \right]^{\alpha_j(t)} 
       \left[ {W^2 - W_{\pi}^2 \over W^2 - W_{\pi}^2 + Q^2 + W_T^2} 
       \right]^{\beta_j(t)}
      \label{eq:g_Delta}
 \ee
where $W_{\pi} \equiv M + m_{\pi}$ is the pion production threshold and $t$ is a parameter aimed at describing the logarithmic scaling  violations in the $DIS$ regime, which we define following Ref. \cite{ALLM} as
 \be
       t = \mbox{ln} \left\{ {\mbox{ln} \left[ (Q^2 + Q_0^2) / \Lambda^2 
       \right] \over \mbox{ln} \left( Q_0^2 / \Lambda^2 \right) } \right\}
      \label{eq:t}
 \ee
In Eq. (\ref{eq:g_Delta}) the parameter $Q_R^2$ describes the transition from the expected dominance of the Regge behaviour at $Q^2 \lsim Q_R^2$
 to the partonic regime at $Q^2 >> Q_R^2$. Indeed, on one hand, for $Q^2 \lsim Q_R^2$ and at large $W^2$ (i.e., low $x$) the r.h.s of Eq. (\ref{eq:g_Delta}) becomes  proportional to $\sum_j a_j W^{2 \alpha_j}$, as expected from the Regge approach. On the other hand, when the Bjorken limit (fixed $x$ and high $Q^2$) is considered, one gets $g^{\Delta \sigma}(x, Q^2) \propto \sum_{j = 1}^N a_j(t) ~ x^{-\alpha_j(t)} (1 - x)^{\beta_j(t)}$. Finally, the quantities $a_j$, $\alpha_j$ and $\beta_j$ are parameters which are assumed to depend linearly on $t$, namely
 \be
       a_j(t) & = & a_j^{(0)} + a_j^{(1)} \cdot t, \nonumber \\
       \alpha_j(t) & = & \alpha_j^{(0)} + \alpha_j^{(1)} \cdot t,
       \nonumber \\
       \beta_j(t) & = & \beta_j^{(0)} + \beta_j^{(1)} \cdot t.
       \label{eq:alpha}
 \ee
 
\indent The parameters appearing in Eqs. (\ref{eq:g_Delta}-\ref{eq:t}) have been determined by fitting existing measurements from Refs. \cite{E143}-\cite{E155} on the asymmetry $A_1(x, Q^2)$ in the $DIS$ kinematics, i.e. outside the resonance regions ($W > 2 ~ GeV$), through the relation
 \be
       A_1(x, Q^2) \to_{DIS} \left( 1 + {4M^2x^2 \over Q^2} \right) 
       {g^{\Delta \sigma}(x, Q^2) \over F_1(x, Q^2)},
       \label{eq:A1_DIS}
 \ee
where $F_1(x, Q^2)$ is the unpolarized proton structure function calculated adopting the interpolation formula developed in Ref. \cite{Ricco}. Following Ref. \cite{ALLM} we have considered $N = 2$ in Eq. (\ref{eq:g_Delta}) fixing the $QCD$ parameter $\Lambda$ at the value $\Lambda = 0.250 ~ GeV$. It turns out that our fitting procedure of $DIS$ data is not very sensitive to the precise value of the parameter $W_T$, since it appears in Eq. (\ref{eq:g_Delta}) only in the combination $Q^2 + W_T^2$. Thus, the value of the parameter $W_T$ is relevant only at low $Q^2$ and, indeed, in the next subsection it will be fixed by requiring the reproduction of the experimental value of the proton $DHG$ sum rule \cite{DHG}; thus, we anticipate here its final value equal to $W_T = 0.475 ~ GeV$. To sum up, we have used $14$ parameters against a total number of experimental points equal to $209$, obtaining for the $\chi^2$ variable divided by the number of d.o.f. the minimum value of $0.66$. The explicit values of our parameters are reported in Table 1, where it can be seen that the transition point from the Regge behaviour to the partonic regime occurs around $Q^2 \simeq Q_R^2 \simeq 4 \div 5 ~ (GeV/c)^2$.

\begin{table}[htb]

{\small {\bf Table 1.} Values of the parameters appearing in Eqs. (\ref{eq:g_Delta}-\ref{eq:alpha})), obtained from the least-$\chi^2$ procedure described in the text.}

\begin{center}

\begin{tabular}{||c|c||c|c||c|c||}
\hline \hline
 $a_1^{(0)}$ & $a_1^{(1)}$ & $\alpha_1^{(0)}$ & $\alpha_1^{(1)}$ & 
 $\beta_1^{(0)}$ & $\beta_1^{(1)}$
\\ \hline \hline
 $~~1.325$ & $~-0.01239$ & $-0.1927$ & $0.2725$ & 
 $0.1250$ & $3.938$
\\ \hline \hline
\end{tabular}

\vspace{0.2cm}

\begin{tabular}{||c|c||c|c||c|c||}
\hline \hline
$a_2^{(0)}$ & $a_2^{(1)}$ & $\alpha_2^{(0)}$ & $\alpha_2^{(1)}$ & 
 $\beta_2^{(0)}$ & $\beta_2^{(1)}$
\\ \hline \hline
 $-2.252$ & $2.099$ & $-0.8718$ & $0.9133$ & 
 $2.910~$ & $3.829$
\\ \hline \hline
\end{tabular}

\vspace{0.2cm}

\begin{tabular}{||c|c||}
\hline \hline
$Q_R^2 ~ (GeV/c)^2$ & $Q_0^2 ~ (GeV/c)^2$
\\ \hline \hline
 $4.498$ & $1.062$
\\ \hline \hline
\end{tabular}

\end{center}

\end{table}

\indent The quality of our fit is shown in Fig. 2. It can be seen that the differences between our fit and the data are approximately distributed as a gaussian-like distribution, and moreover they almost do not exceed the statistical + systematic errors of the data (added in quadrature). The uncertainty on our parameterization of $g^{\Delta \sigma}$ generated by the fitting procedure has been estimated through the uncertainties obtained for the values of the parameters reported in Table 1 from our least-$\chi^2$ procedure. Using different bins in the variable $x$ we have found that the uncertainty on our $g^{\Delta \sigma}$ can be approximated by the following simple formula: $(1 - 0.5 \cdot x / x_{\pi}) \cdot 15\%$, where $x_{\pi} = Q^2 / (Q^2 + W_{\pi}^2 - M^2)$ is the pion threshold in $x$.

\begin{figure}[htb]

\centerline{\epsfxsize=16cm \epsfig{file=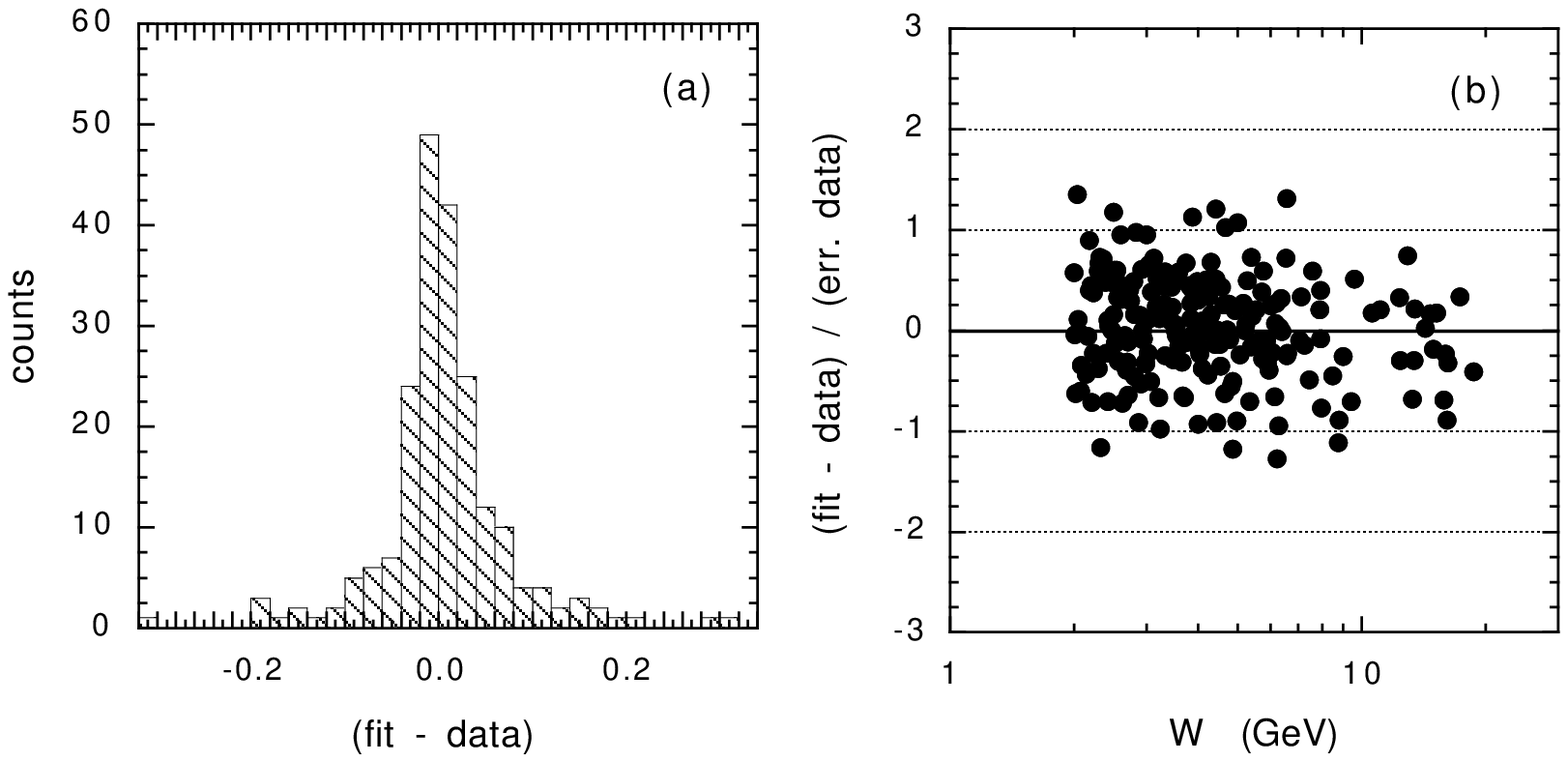}}

\small{ {\bf Figure 2}. (a) Distribution of the differences between our fit (\ref{eq:A1_DIS}) and existing data \cite{E143}-\cite{E155} on the asymmetry $A_1(x, Q^2)$ in $DIS$ kinematics only ($W > 2 ~ GeV$).  (b) Ratio of the differences in (a) with the statistical + systematic errors of the data (added in quadrature) versus the final invariant mass $W$.}

\end{figure}

\indent We point out that in Eq. (\ref{eq:g_Delta}) $g^{\Delta \sigma}$ is assumed to behave in the Bjorken limit as a power of $x$ at low values of $x$. Since there is no strong argument in favor of such an assumption (cf., e.g., the discussion of the limit $x \to 0$ in Refs. \cite{Altarelli,Reya}), Eq. (\ref{eq:g_Delta}) has to be considered as a simple approximation valid in a limited $x$-range. In this respect we have already observed from Fig. 1 that existing data are scarce below $x \sim 0.02$; thus, we consider $x \gsim 0.02$ as the $x$-range of applicability of our parameterization (\ref{eq:g_Delta}), which is clearly well enough for our main purpose to study leading and higher twist effects at large-$x$.

\indent The contribution of $g^{LT}(x, Q^2)$ to Eq. (\ref{eq:g12_NR}) is marginal in $DIS$ kinematics and, moreover, the whole effect of $g_2(x, Q^2)$ in the Nachtmann moments (\ref{eq:M1}) is power suppressed; therefore, we do not need a very refined interpolation formula for $g_2(x, Q^2)$. In this respect the analyses of the $SMC$ \cite{SMC}, $E143$ \cite{E143} and $E155$ \cite{E155_g2} $DIS$ data on $A_2(x, Q^2)$ suggest that $g_2(x, Q^2)$ is consistent with the Wandzura-Wilczek relation \cite{WW}. Thus, we impose a simple-minded generalization of the latter as a constraint on our parameterization for $g^{LT}(x, Q^2)$; we simply assume that
 \be
       g^{LT}(x, Q^2) = F_{thr} \int_x^{x_{\pi}} dx' {g^{\Delta \sigma}(x', 
       Q^2) \over x'}
       \label{eq:g_LT}
 \ee
where $F_{thr} = \sqrt{1 - W_{\pi}^2 / W^2}$ is a threshold factor. Thus, our interpolation model for the non-resonant contributions to $g_i(x, Q^2)$ reads as
 \be
       \label{eq:g1_non-res}
       g_1^{(non-res.)}(x, Q^2) & = & g^{\Delta \sigma}(x, Q^2) + {4 M^2 x^2 
       \over Q^2} F_{thr} \int_x^{x_{\pi}} dx' {g^{\Delta \sigma}(x', Q^2) 
       \over x'} \\
       \label{eq:g2_non-res}
       g_2^{(non-res.)}(x, Q^2) & = & - g^{\Delta \sigma}(x, Q^2) + F_{thr} 
       \int_x^{x_{\pi}} dx' {g^{\Delta \sigma}(x', Q^2) \over x'}
 \ee
The corresponding predictions for $A_2(x, Q^2) \equiv 2 M x \cdot [g_1(x, Q^2) + g_2(x, Q^2)] / [Q \cdot F_1(x, Q^2)]$ compare positively against available $DIS$ data, as shown in Fig. 3.

\begin{figure}[htb]

\centerline{\epsfxsize=14cm \epsfig{file=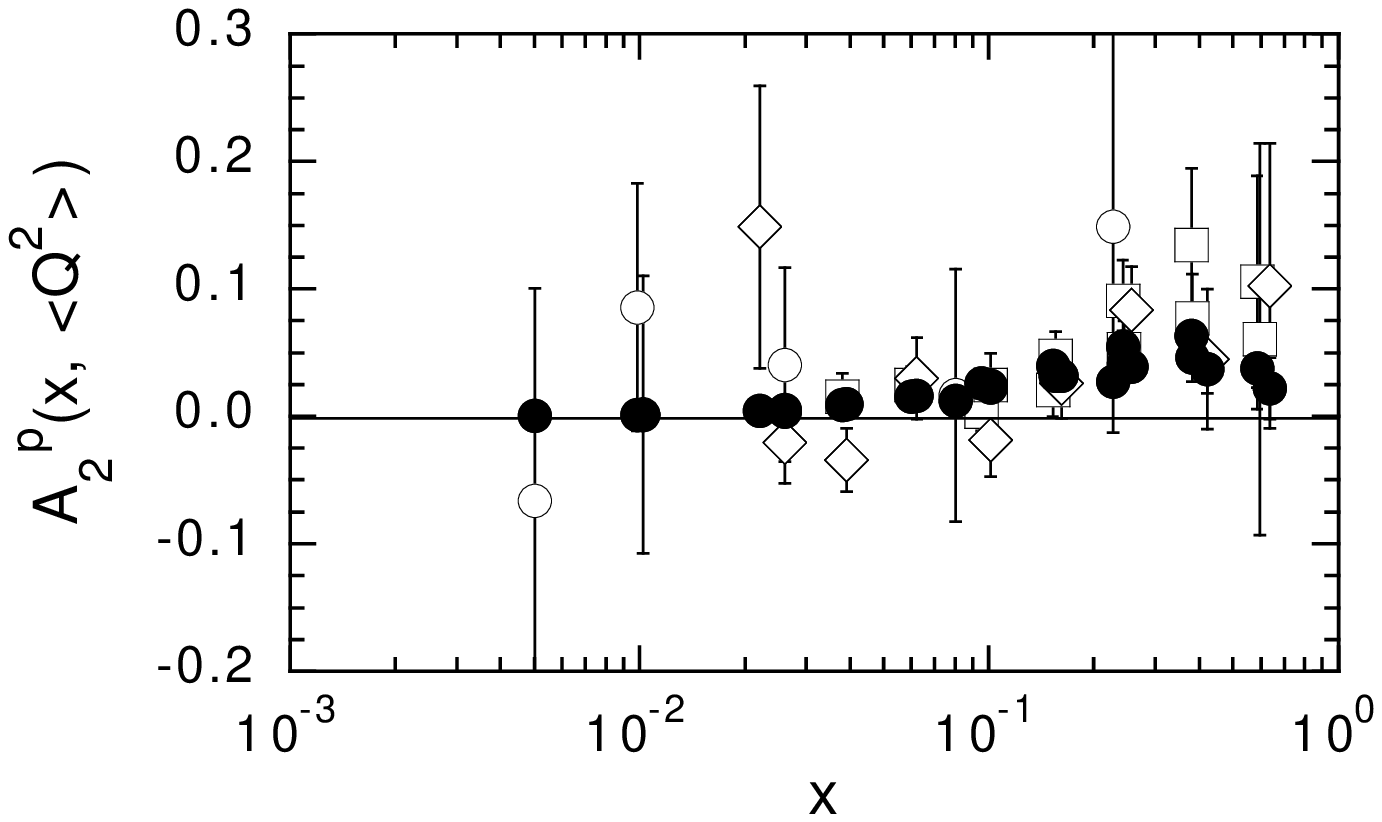}}

\small{ {\bf Figure 3}. Values of the asymmetry $A_2^p(x, \langle Q^2 \rangle)$ for the proton versus the Bjorken variable $x$ in case of $DIS$ kinematics ($W > 2 ~ GeV$).  The open dots, squares and diamonds correspond to the data of the $SMC$ \protect\cite{SMC}, $E143$ \protect\cite{E143} and $E155$ \protect\cite{E155_g2} experiments, respectively. The full dots are our predictions based on Eqs. (\protect\ref{eq:g1_non-res} - \protect\ref{eq:g2_non-res}). The values of $\langle Q^2 \rangle$ are those corresponding to the kinematics of the various experiments for each value of $x$.}

\end{figure}

\subsection{Resonant contribution and the $DHG$ sum rule}

\indent In the nucleon-resonance regions ($W \lsim 2 ~ GeV$) we follow the approach of Refs. \cite{Walker,Weise} by adopting a simple Breit-Wigner shape to describe the $W$-dependence of the contribution of an isolated resonance $R$, while its $Q^2$-dependence can be conveniently expressed in terms of transverse helicity amplitudes $A_{1/2}^R$ and $A_{3/2}^R$ as well as the longitudinal helicity amplitude $S_{1/2}^R$, viz.
 \be
       \label{eq:g1_res}
       g_1^{(res.)}(x, Q^2) & = & {M K \over 4 \pi^2 \alpha_{em}} {1 \over 1 
       + 4 M^2 x^2 / Q^2} \left[ {\sigma_{1/2}^{(res.)} - 
       \sigma_{3/2}^{(res.)} \over 2} + {2 M x \over Q} \sigma_{LT}^{(res.)} 
       \right]  \\
       \label{eq:g2_res}
       g_2^{(res.)}(x, Q^2) & = & {M K \over 4 \pi^2 \alpha_{em}} {1 \over 1 
       + 4 M^2 x^2 / Q^2} \left[ - {\sigma_{1/2}^{(res.)} - 
       \sigma_{3/2}^{(res.)} \over 2} + {Q \over 2 M x} \sigma_{LT}^{(res.)} 
       \right] 
 \ee
with
 \be
        \label{eq:sigmaT}
        \sigma_{1/2 (3/2)}^{(res.)} & = & \sum_{R} {q_R^2 \over q_W^2} {4 M 
        M_R \Gamma_R \over (W^2 - M_R^2)^2 + M_R^2 \Gamma_R^2} 
        |A_{1/2 (3/2)}^R|^2 \\
        \label{eq:sigmaLT}
        \sigma_{LT}^{(res.)} & = & \sum_{R} {q_R^2 \over q_W^2}  {4 M M_R 
        \Gamma_R \over (W^2 - M_R^2)^2 + M_R^2 \Gamma_R^2} {Q \over 
        \sqrt{2} q_{cm}} \left( S_{1/2}^R \right)^* A_{1/2}^R \\
        \label{eq:GammaR}
        \Gamma_R & = & \Gamma_R^{(0)} \left( {q_W \over q_R} \right)^{2 
        \ell_R + 1} \left( {q_R^2 + X_R^2 \over q_W^2 + X_R^2} 
        \right)^{\ell_R}
 \ee
where $M_R$ and $\Gamma_R^{(0)} \equiv \Gamma_R(W = M_R)$ are the mass and the width of the resonance $R$, respectively, and eventually $q_W \equiv \sqrt{(W^2 + M^2 - m_{\pi}^2) / 4W^2 - M^2}$, $q_R \equiv q_W(W = M_R)$ and $q_{cm} = \sqrt{Q^2 + (W^2 - M^2 - Q^2)^2 / 4 W^2}$. The parameters $\ell_R$ and $X_R$ have the same meaning as in Ref. \cite{Walker}.

\indent We point out that the $W$-shape encoded in Eqs. (\ref{eq:sigmaT}-\ref{eq:GammaR}) is inspired by the results of Ref. \cite{Walker}, and therefore it differs from the one adopted in Ref. \cite{Weise}, mainly because the width parameter $\Gamma_R$ is assumed to be a constant independent on $W$ in \cite{Weise} at variance with Eq. (\ref{eq:GammaR}). The use of the latter however allows to achieve a proper reproduction of the asymmetric shape around the resonance bumps due to $W$-dependence of the available phase space for resonance decays. Such a dependence produces an important shift of the location of the resonance peaks with respect to $W = M_R$, which is neglected in Ref. \cite{Weise}.

\indent To develop our interpolation formula in the resonance regions we have included in Eqs. (\ref{eq:sigmaT}-\ref{eq:sigmaLT}) all the "four-star" resonances of the $PDG$ \cite{PDG} having a mass $M_R < 2 ~ GeV$ and a total transverse amplitude $\sqrt{|A_{1/2}^R|^2 + |A_{3/2}^R|^2}$ larger than $0.050 ~ GeV^{-1/2}$ at the photon point. Following Ref. \cite{Weise} we have represented the $Q^2$ behavior of the transverse helicity amplitudes in the following form
 \be
       |A_{1/2, 3/2}^R| = \sqrt{ {1 \pm A_1^R(Q^2) \over 2}} C_R e^{- B_R 
       \cdot Q^2},
       \label{eq:AT}
 \ee
while for the $LT$ cross section we have introduced the parameter $A_{LT}^R$ defined as 
 \be
     A_{LT}^R \equiv {S_{1/2}^R \over A_{1/2}^R}
     \label{eq:ALT}
 \ee
so that  $\sigma_{LT}^R = A_{LT}^R \cdot (Q / \sqrt{2} q_{cm}) \sigma_{1/2}^R$. Note that in Eq. (\ref{eq:AT}) the parameter $C_R$ represents just the total transverse amplitude $\sqrt{|A_{1/2}^R|^2 + |A_{3/2}^R|^2}$ at the photon point. Finally, as for the background under the resonances we adopt directly  Eq. (\ref{eq:g1_non-res}) without any adjustment of the parameters, thanks to the fact that the latter smoothly behaves for $W < 2 ~ GeV$.

\indent Firstly we have taken the values of the widths $\Gamma_R^{(0)}$, of the amplitudes $C_R$ and of the asymmetry $A_1^R$ at the photon point from $PDG$ \cite{PDG}. We have then calculated  the asymmetry of the transverse cross section $\sigma_{3/2} - \sigma_{1/2}$ as a function of the photon energy $E_{\gamma} = (W^2 - M^2) / 2M$, and compared our results with  the recent data from Mainz \cite{Mainz}. It turns out that an adjustment of the widths $\Gamma_{\Delta(1232)}^{(0)}$ and $\Gamma_{D_{13}(1520)}^{(0)}$ as well as of the amplitude $C_{D_{13}(1520)}$ is needed. Instead of their $PDG$ values (put in parenthesis) we adopt: $\Gamma_{\Delta(1232)}^{(0)} = 0.10 ~ (0.12) ~ GeV$, $\Gamma_{D_{13}(1520)}^{(0)} = 0.20 ~ (0.13) ~ GeV$ and $C_{D_{13}(1520)} = 0.200 ~ (0.170) ~ GeV^{-1/2}$. Finally, as anticipated in the previous subsection, the value of the parameter $W_T$ appearing in Eq. (\ref{eq:g_Delta}) is fixed at the value $W_T = 0.475 ~ GeV$ in order to reproduce the experimental value of proton $DHG$ sum rule \cite{DHG}, viz. 
 \be
       \int_{E_{\pi}}^{\infty} dE_{\gamma} {\sigma_{3/2} - \sigma_{1/2} 
       \over E_{\gamma}} = {2 \pi^2 \alpha_{em} \kappa^2 \over M^2} \simeq 
       204.5 ~ \mu barn
       \label{eq:DHG}
 \ee
where $E_{\pi}$ is the pion threshold in terms of the photon energy $E_{\gamma}$ and $\kappa$ is the (proton) anomalous magnetic moment. Our final results at $Q^2 = 0$ are reported in Fig. 4 and positively compared with the Mainz data \cite{Mainz}. It can clearly be seen that our parameterization of the non-resonant term (\ref{eq:g1_non-res}) obtained from fitting $DIS$ data, can be safely extended down to the photon point to describe a smooth background under the resonance bumps.

\begin{figure}[htb]

\centerline{\epsfxsize=14cm \epsfig{file=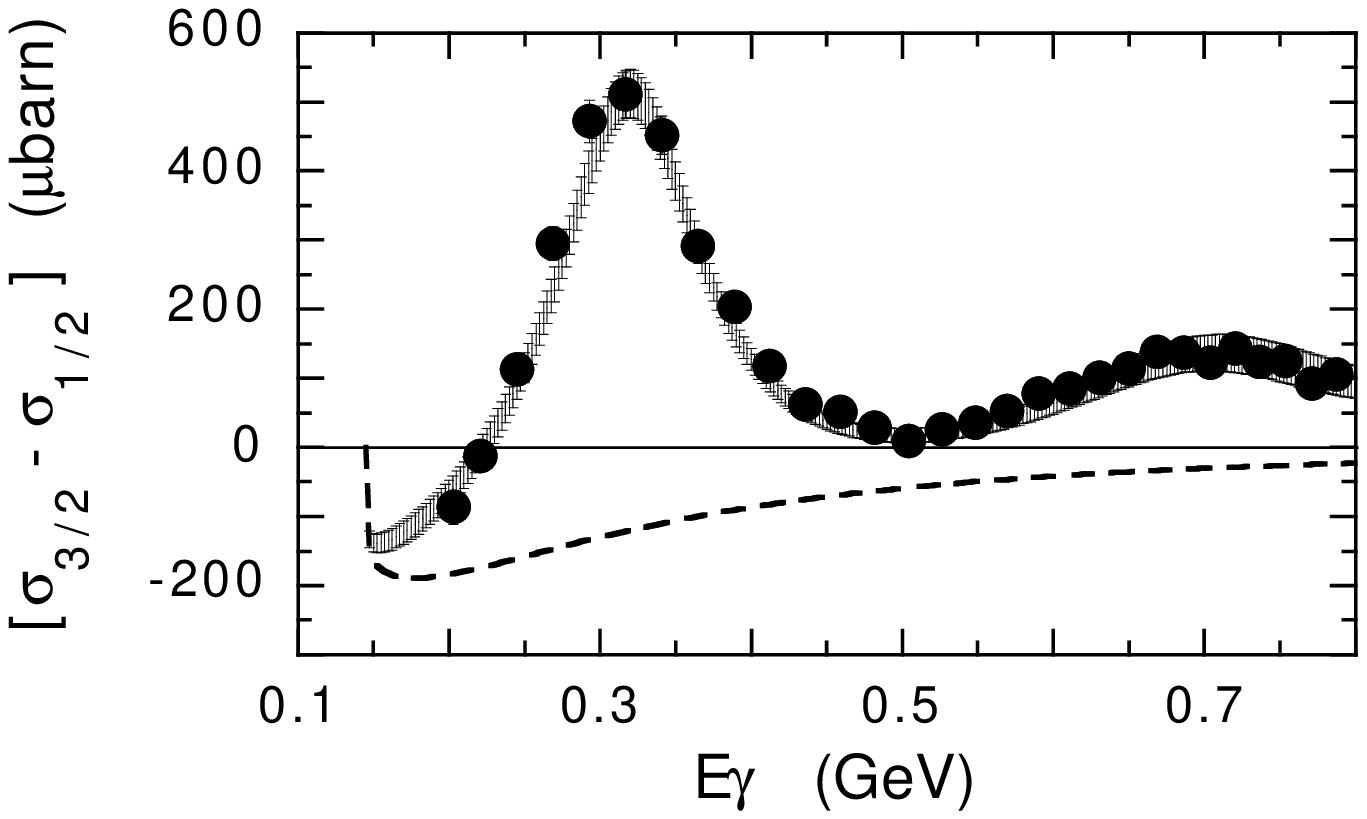}}

\small{ {\bf Figure 4}. Asymmetry of the proton transverse cross sections, [$\sigma_{3/2} - \sigma_{1/2}$], versus the photon energy $E_{\gamma}$. Full dots are the experimental data from Ref. \cite{Mainz}. The shaded area is our prediction, as explained in the text, while the dashed line is our non-resonant contribution.}

\end{figure}

\indent In Table 2 we have reported the contribution of various integration regions over the photon energy $E_{\gamma}$ to the proton $DHG$ sum rule (\ref{eq:DHG}) and to the forward spin polarizability $\gamma_0$, defined as
 \be
       \gamma_0 = -{1 \over 4 \pi^2} \int_{E_{\pi}}^{\infty} dE_{\gamma} 
       {\sigma_{3/2} - \sigma_{1/2} \over E_{\gamma}^3}
       \label{eq:spin}
 \ee
It can be seen that an important contribution to both the $DHG$ sum rule and the forward spin polarizability comes from the experimentally unobserved region below $E_{\gamma} = 0.2 ~ GeV$ in overall agreement with the prediction of the Unitarity Isobar Model of Ref. \cite{UIM}, while the contribution of our high-energy tail above $E_{\gamma} = 1.6 ~ GeV$ to the $DHG$ sum rule is just half of the prediction of Ref. \cite{Bianchi}. Note that our parameterization of $g^{\Delta \sigma}$ (\ref{eq:g_Delta}) implies that at $Q^2 = 0$ the transverse cross section asymmetry $[\sigma_{3/2} - \sigma_{1/2}]$ behaves as $E_{\gamma}^{\alpha_j(0) - 1}$ at high photon energies. The negative valus obtained for the parameters $\alpha_j^{(0)}$ (see Table 1) largely ensure that the $DHG$ integral (\ref{eq:DHG}) is convergent at high energies and does not require subtractions.

\begin{table}[htb]

{\small {\bf Table 2.} Contributions of various integration regions over the photon energy $E_{\gamma}$ (in $GeV$) to the proton $DHG$ sum rule (\ref{eq:DHG}) and to the forward spin polarizability $\gamma_0$ (\ref{eq:spin}), calculated using our interpolation formula.}

\begin{center}

\begin{tabular}{||c||c|c||}
\hline \hline
$integration ~ region$ & $DHG ~ (\mu barn)$ & $\gamma_0 ~ (10^{-6} ~ fm^4)$
\\ \hline \hline
$E_{\gamma} \leq 0.2$ & $-37 \pm 5$ & $133 \pm 17$ \\ \hline
$0.2 \leq E_{\gamma} \leq 0.8$ & $ 207 \pm 20$ & $-173 \pm 10$ \\ \hline
$0.8 \leq E_{\gamma} \leq 1.6$ & $ 47 \pm 9$ & $-5 \pm 1$ \\ \hline
$E_{\gamma} \geq 1.6$ & $ -13 \pm 2$ & $< 0.1$ \\ \hline \hline
$total$ & $204 \pm 23$ & $-45 \pm 20$ \\ \hline \hline
\end{tabular}

\end{center}

\end{table}

\indent The $Q^2$ dependence of $A_1^R$ and the values of the parameter $B_R$ and $A_{LT}^R$ have been estimated using available electroproduction data from Ref. \cite{Burkert}. The uncertainty on our parameterization in the resonance regions has been estimated by assigning to the slope parameter $B_R$ an overall $25 \%$ uncertainty and to the parameter $C_R$ the uncertainty $\Delta C_R$ arising from the reported $PDG$ uncertainties on the transverse helicity amplitudes at the photon point. Table 3 collects the values of all the relevant parameters necessary for evaluating Eqs. (\ref{eq:sigmaT}-\ref{eq:GammaR}). We have to mention that: ~ i) the values adopted for the parameters $A_{LT}^R$ fully satisfy the bounds imposed by the existing data on the unpolarized longitudinal to transverse ratio (see Ref. \cite{Thia}), and ~ ii) our parameterization of the full structure function $g_2(x, Q^2)$ is consistent with the Burkhardt-Cottingham sum rule \cite{BCS}, stating that $\int_0^1 dx ~ g_2(x, Q^2) = 0$, within $2 \sigma$ standard deviations at all values of $Q^2$.

\begin{table}[htb]

{\small {\bf Table 3.} Values of the parameters appearing in Eqs. (\ref{eq:sigmaT}-\ref{eq:GammaR}) and obtained as described in the text. Resonance masses are in $MeV$, $C_R$ in $GeV^{-1/2}$, $\Delta C_R$ in $\%$, $B_R$ in $GeV^{-2}$, $\Gamma_R^{(0)}$, $X_R$ and $Q$ in $GeV$.}

\begin{center}

\begin{tabular}{||c||c|c|c|c|c|c|c|c||}
\hline \hline
 $resonance$ & $C_R$ & $\Delta C_R $ & $A_1^R$ & $B_R$ & $\Gamma_R^{(0)}$ & 
 $A_{LT}^R$ & $\ell_R$ & $X_R$
\\ \hline \hline
 $\Delta (1232)$ & $0.290$ & $5$ & $-0.56$ & $0.7$ & $0.10$ & $-0.1$ & $1$ &
 $0.16$
\\ \hline
 $P_{11} (1440)$ & $0.065$ & $5$ & $~1.0$ & $1.6$ & $0.30$ & $-0.2$ & $1$ &
 $0.35$
\\ \hline
 $D_{13} (1520)$ & $0.200$ & $10$ & $1-e^{0.67-0.65 Q}$ & $0.8$ & $0.20$ &
 $~~0.2$ & $2$ & $0.35$
\\ \hline
 $S_{11} (1535)$ & $0.090$ & $35$ & $~1.0$ & $0.6$ & $0.15$ & $~~0.2$ & $1$
 & $0.35$
\\ \hline
 $S_{11}^* (1650)$ & $0.055$ & $30$ & $~1.0$ & $1.0$ & $0.15$ & $-0.3$ & $1$
 & $0.35$
\\ \hline
 $D_{15}+F_{15} (1680)$ & $0.150$ & $15$ & $1-e^{0.69-0.98 Q}$ & $0.6$ &
 $0.13$ & $-0.3$ & $3$ & $0.35$
\\ \hline
 $D_{33} (1700)$ & $0.135$ & $30$ & $~0.2$ & $1.0$ & $0.30$ & $-0.3$ & $2$ &
 $0.35$
\\ \hline
 $F_{35} (1905)$ & $0.055$ & $60$ & $-0.50$ & $0.6$ & $0.25$ & $-0.3$ & $3$
 & $0.35$
\\ \hline
 $F_{37} (1950)$ & $0.125$ & $20$ & $-0.24$ & $0.6$ & $0.25$ & $-0.3$ & $3$
 & $0.35$
\\ \hline \hline
\end{tabular}

\end{center}

\end{table}

\indent The quality of our interpolation formula in the resonance electroproduction regions is illustrated in Fig. 5, where our predictions are positively compared with the two sets of existing data from $E143$ \cite{E143} at $Q^2 \sim 0.5$ and $\sim 1.2 ~ (GeV/c)^2$.

\begin{figure}[htb]

\centerline{\epsfxsize=16cm \epsfig{file=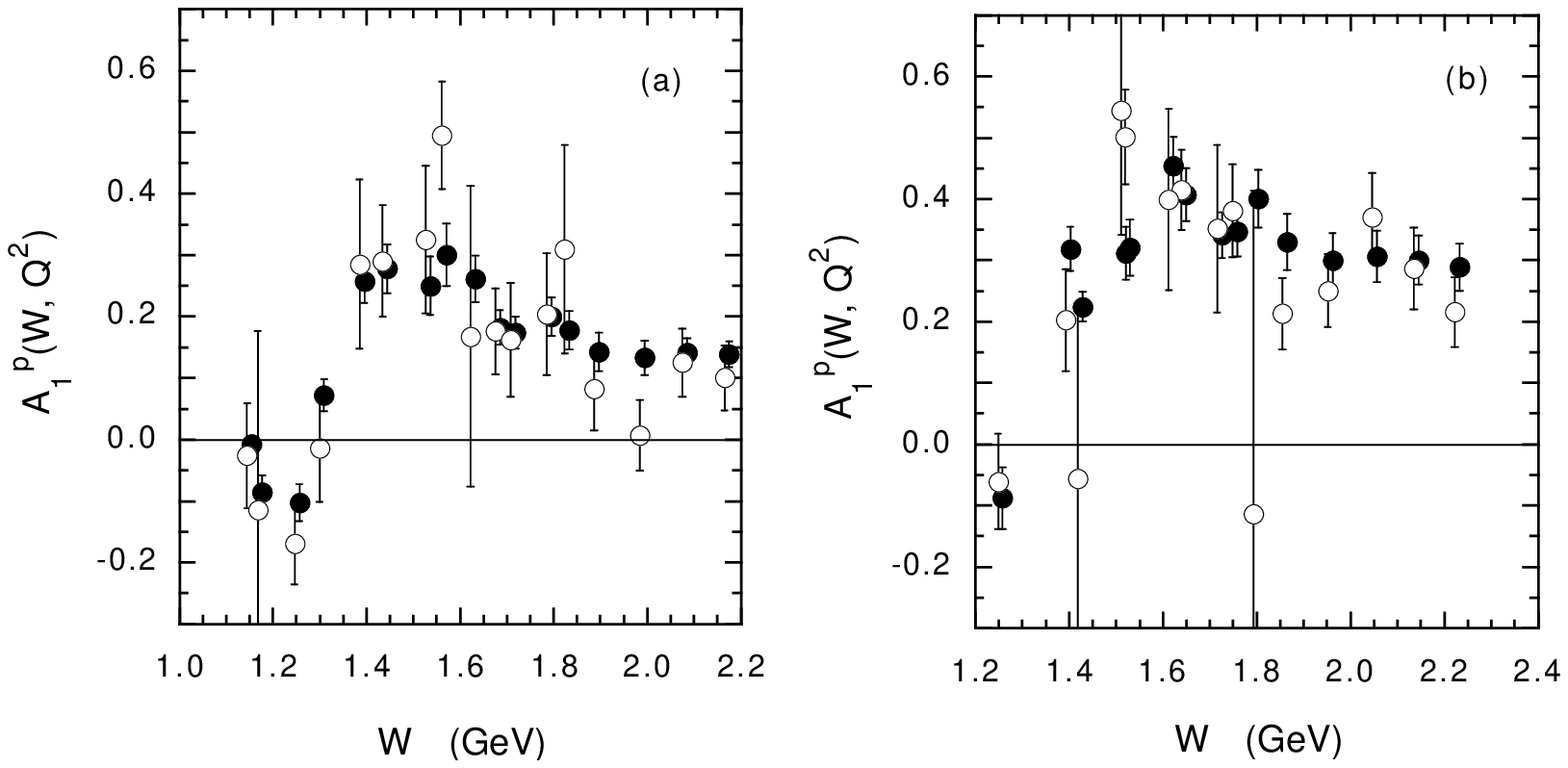}}

\small{ {\bf Figure 5}. Comparison of the proton transverse asymmetry $A_1^p(W, Q^2)$ obtained through our interpolation formula (full dots) with the existing data from Ref. \cite{E143} (open dots) in the resonance regions. The average value of $Q^2$ is $\sim 0.5$ and $\sim 1.2 ~ (GeV/c)^2$ in (a) and (b), respectively.}

\end{figure}

\indent Before going on with the calculation of the Nachtmann moments (\ref{eq:M1}), our prediction for the generalized $DHG$ integral, defined as
 \be
       I_{DHG}(Q^2) = {2M^2 \over Q^2} \int_0^{x_{\pi}} dx ~ g_1(x, Q^2) ~,
       \label{eq:DHG_gen}
 \ee
is presented in Fig. 6 and positively compared with available experimental data. Note that in Eq. (\ref{eq:DHG_gen}) the upper limit of integration excludes the elastic contribution and therefore the generalized $DHG$ integral $I_{DHG}(Q^2)$ cannot be analyzed in terms of twists, because the $OPE$ is fully inclusive.

\indent As it is well known, at the photon point the (proton) generalized $DHG$ integral
 \be
       I_{DHG}(Q^2 = 0) = {M^2 \over 8 \pi^2 \alpha_{em}} 
       \int_{E_{\pi}}^{\infty} dE_{\gamma} {\sigma_{1/2} - \sigma_{3/2} 
       \over E_{\gamma}} = -{\kappa^2 \over 4}
       \label{eq:IDHG}
 \ee
is negative ($\simeq -0.80$), while it becomes positive in the $DIS$ regime. Therefore it should cross zero at some value of $Q^2$ and according to the predictions of our parameterization (see Fig. 6) the zero-crossing point is expected to occur at $Q^2 = Q_X^2= 0.16 \pm 0.04 ~ (GeV/c)^2$. Such a value is significantly below the prediction $Q_X^2 \sim 0.8 ~ (GeV/c)^2$ found in Ref. \cite{BI}, where the dominance of the $N - \Delta(1232)$ transition in Eq. (\ref{eq:DHG_gen}) at low $Q^2$ was assumed and a vector meson dominance picture was adopted for the non-resonant background. The difference is due to the significative contribution of non-resonant processes present at low $Q^2$ in our parameterization (see dashed curve in Fig. 6) at variance with the assumptions made in Ref. \cite{BI}. Our zero-crossing point is below the finding $Q_X^2 \sim 0.3 ~ (GeV/c)^2$ obtained in Ref. \cite{Ma} within the constituent quark model and characterized by the so-called dominance of low-lying resonances at low $Q^2$. Our result is quite close to the prediction $Q_X^2 \sim 0.2 ~ (GeV/c)^2$ of Ref. \cite{ST}, where the origin of the zero-crossing point is traced back to the strong $Q^2$-dependence of the first moment of the inelastic part of the structure function $g_2(x, Q^2)$ implied by the Burkhardt-Cottingham sum rule \cite{BCS}. Therefore, our finding, if confirmed by a direct measurement of the $Q^2$-behavior of the generalized $DHG$ integral, should provide an important constraint on hadronic models and on the physics of nucleon resonances.

\begin{figure}[htb]

\centerline{\epsfxsize=14cm \epsfig{file=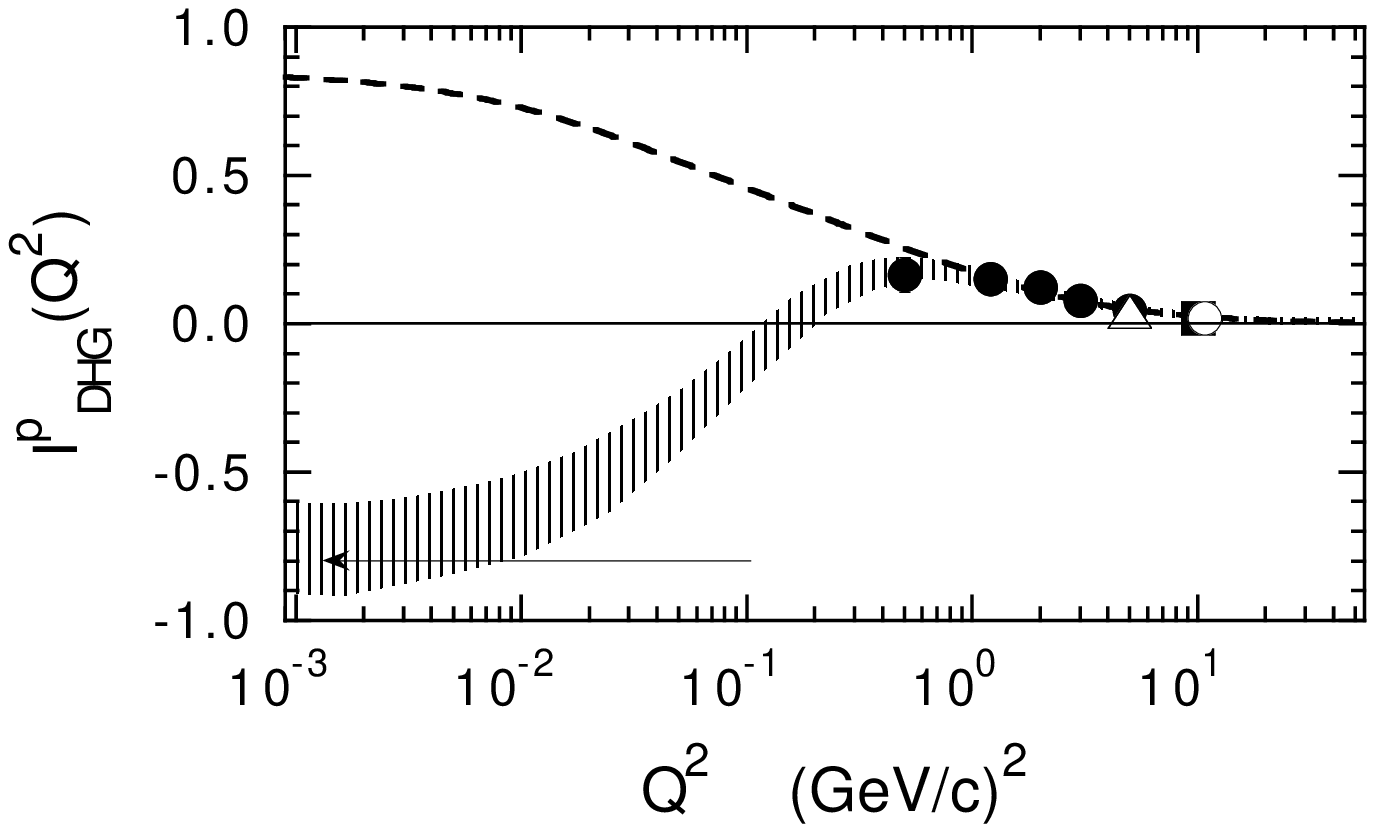}}

\small{ {\bf Figure 6}. Generalized $DHG$ integral (\ref{eq:DHG_gen}) for the proton versus $Q^2$. Full dots and squares, open dots and triangles are the experimental results of Refs. \cite{E143,SMC,EMC,E155}. The shaded area is the prediction based on our parameterization of $g_1^p(x, Q^2)$, while the dashed line corresponds to the non-resonant contribution only. The arrows indicates the location of the value of $DHG$ sum rule ($\simeq -0.80$) at the photon point.}

\end{figure}

\indent The Nachtmann moments $M_n^{(1)}(Q^2)$ [see Eq. (\ref{eq:M1})] can now be evaluated through our interpolation formulae for the polarized structure functions $g_1(x, Q^2)$ and $g_2(x, Q^2)$, which we stress have the following range of applicability: $x \gsim 0.02$ and $Q^2  \lsim 50 ~ (GeV/c)^2$. The separate contributions of the elastic peak (\ref{eq:g1_el}-\ref{eq:g2_el}), the resonances (\ref{eq:g1_res}-\ref{eq:sigmaLT}) and the inelastic contribution [i.e. the sum of the resonant and non-resonant parts (\ref{eq:g1_non-res}-\ref{eq:g2_non-res})] are shown in Fig. 7 for $0.1 \leq Q^2 ~ (GeV/c)^2 \leq 50$. It can be seen that the elastic term is dominant for $Q^2 \lsim n/2 ~ (GeV/c)^2$, while the inelastic contribution exhausts the Nachtmann moment for $Q^2 \gsim n ~ (GeV/c)^2$. Note that, at variance with the generalized $DHG$ integral (\ref{eq:DHG_gen}), the first moment $M_1^{(1)}(Q^2)$ does not become negative at low values of $Q^2$ because of the elastic contribution (\ref{eq:M1_el}), as firstly pointed out in Ref. \cite{Wally}.  Moreover, the inelastic parts of the Nachtmann moments shown in Fig 7(b-d) change their sign at $Q^2 \simeq 0.45 \pm 0.10, ~ 0.75 \pm 0.15, ~ 1.1 \pm 0.2 ~ (GeV/c)^2$ for $n = 3, ~ 5, ~ 7$, respectively.

\begin{figure}[htb]

\centerline{\epsfxsize=16cm \epsfig{file=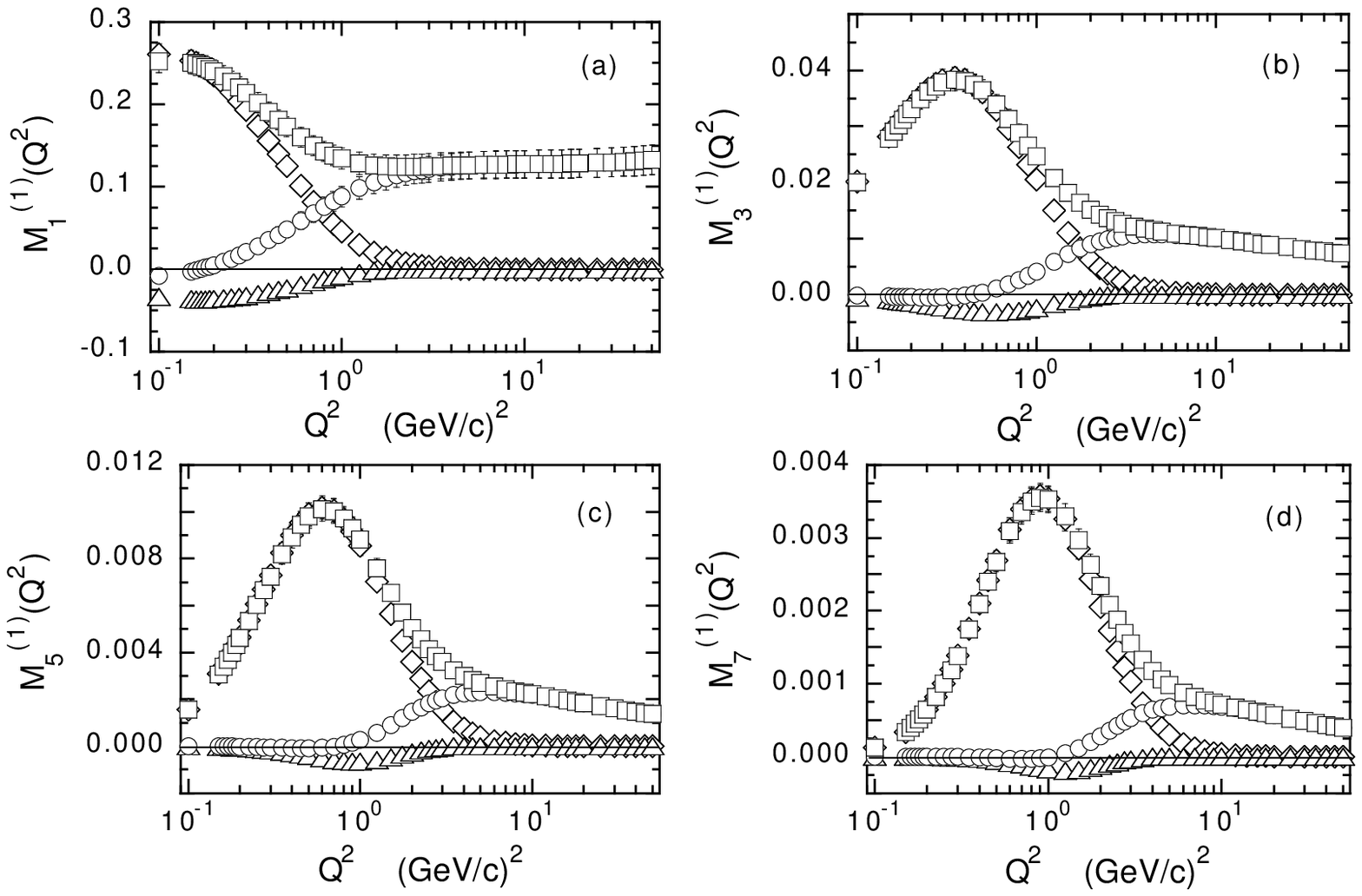}}

\small{ {\bf Figure 7}. Proton Nachtmann moments $M_n^{(1)}(Q^2)$ [see Eq. (\ref{eq:M1})] versus $Q^2$ for $n = 1$ (a), $n = 3$ (b), $n = 5$ (c) and $n = 7$ (d). Open diamonds and triangles correspond to the contributions of the elastic peak (\ref{eq:g1_el}-\ref{eq:g2_el}) and the resonances (\ref{eq:g1_res}-\ref{eq:sigmaLT}), respectively. Open dots are the inelastic contribution, i.e. the sum of the resonant and non-resonant parts (\ref{eq:g1_non-res}-\ref{eq:g2_non-res}). Open squares represent the full Nachtmann moment given by the sum of the elastic and inelastic parts.}

\end{figure}

\indent Finally, we want to mention that quite recently \cite{Klaus} preliminary photoabsorption data above $E_{\gamma} \simeq 0.8 ~ GeV$ have become available from a $GDH$ experiment at $ELSA$. Such data cover a large kinematical region where the $F_{15}(1680)$ nucleon resonance contributes predominantly. It appears that our fit, based on the $PDG$ values for the helicity amplitudes at the photon point, overestimates the preliminary data. The agreement can be recovered simply by imposing a $\simeq 25 \%$ reduction of the strength parameter $C_{F_{15}(1680)}$ and a $\simeq 10 \%$ reduction of the parameter $W_T$ in order to keep the $GDH$ sum rule fulfilled. We do not include these modifications in our present parameterization, leaving this issue to be fixed when final data from $ELSA$ will be available. Nevertheless, we have checked that the above-mentioned changes do not modify within the quoted uncertainties both the zero-crossing point of the generalized $GDH$ integral and the twist analysis of the next two Sections.

\subsection{Bloom-Gilman local duality}

\indent An important feature of the results shown in Fig. 7 is that the resonant contribution is negative for $Q^2 \sim$ few $(GeV/c)^2$. This is mainly due to the well established fact \cite{Burkert} that the proton transverse asymmetry $A_1^p$ in the regions of  the $\Delta(1232)$-resonance electroproduction is negative up to $Q^2 \sim 3 \div 4 ~ (GeV/c)^2$. Thus, for $Q^2 \sim$ few $(GeV/c)^2$ the resonant contribution to the polarized proton structure function is opposite in sign with respect to the unpolarized case (see Ref. \cite{Ricco}). In this respect we want ot remind that the concept of parton-hadron local duality makes an important (and not yet fully understood) connection between the physics in the nucleon-resonance and $DIS$ regions. Indeed, the parton-hadron local duality, observed empirically \cite{BG} by Bloom and Gilman in the unpolarized transverse structure function of the proton, states that the smooth scaling curve measured in the $DIS$ region at high $Q^2$ represents an average over the resonance bumps seen in the same $x$ region at low $Q^2$. More precisely, it occurs a precocious scaling of the average of the $F_2^p(\xi, Q^2)$ data in the resonance regions to the $DIS$ structure function $F_2^p(\xi)$, at corresponding values of the improved scaling variable $\xi$ \cite{TM}.

\begin{figure}[htb]

\centerline{\epsfxsize=16cm \epsfig{file=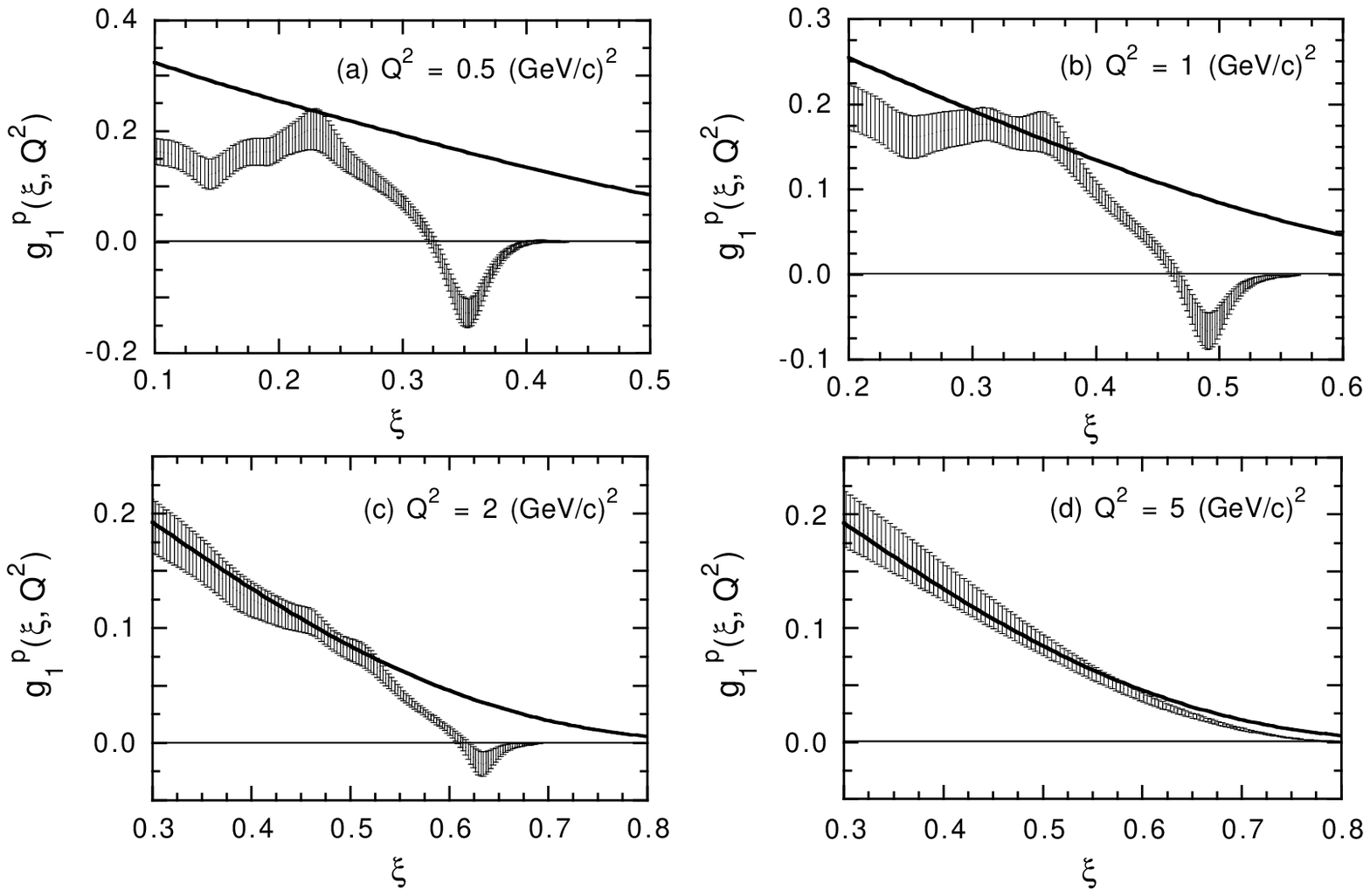}}

\small{ {\bf Figure 8}. Polarized proton structure function $g_1^p(\xi, Q^2)$ versus the Nachtmann variable $\xi$ for various values of $Q^2$. In (a)-(d) the shaded areas are pseudo-data generated through our interpolation formula at $Q^2 = 0.5, 1, 2$ and $5 ~ (GeV/c)^2$, respectively.  The solid line is the result of our parameterization evaluated at $Q^2 = 20 ~ (GeV/c)^2$.}

\end{figure}

\indent It is therefore legitimate to ask ourselves whether the $BG$ local duality holds as well in the polarized case. To this end we have generated pseudo-data in the resonance regions and in the $DIS$ regime via our interpolation formula for $g_1^p(x, Q^2)$. Our results are reported in Fig. 8, where it can clearly be seen that: ~ i) at values of $Q^2$ as low as $\sim 0.5 ~ (GeV/c)^2$ there is no evidence at all of an occurrence of the $BG$ local duality, as in the case of the unpolarized transverse structure function of the proton (see Ref. \cite{duality})), and ~ ii)  in the kinematical regions where the $\Delta(1232)$ resonance is prominently produced, the $BG$ local duality breaks down at least for $Q^2$ up to few $(GeV/c)^2$, while in the higher resonance regions for $Q^2 \gsim 1 ~ (GeV/c)^2$ it is not excluded by our parameterization. Note that in the unpolarized case the onset of the $BG$ local duality occurs at $Q^2 \simeq 1 \div 2 ~ (GeV/c)^2$ \cite{BG,duality}, including also the $\Delta(1232)$ resonance regions \cite{Carlson}. It should be mentioned that the usefulness of the concept of local duality relies mainly on the possibility to address the $DIS$ curve at large $\xi$ through measurements at low $Q^2$ in the resonance regions. It is therefore clear that the breakdown of the local duality in the region of the $\Delta(1232)$ resonance forbid us to get information from duality on the behavior of the scaling curve at the highest values of $\xi$.

\section{$NLO$ analysis of the polarized Nachtmann moments}

\indent  In this Section we present our power correction analysis of the polarized Nachtmann moments (\ref{eq:M1}), adopting for the leading twist the $NLO$ approximation. Following Refs. \cite{Ricco,SIM00} a phenomenological ansatz is introduced for describing power corrections, viz.
 \be
       M_n^{(1)}(Q^2) = \delta\mu_n^{(1)}(Q^2) + \delta a_n^{(4)} ~ \left[ 
       {\alpha_s(Q^2) \over \alpha_s(\mu^2)} \right]^{\delta \gamma_n^{(4)}} 
       {\mu^2 \over Q^2} +  \delta a_n^{(6)} ~ \left[ {\alpha_s(Q^2) \over 
       \alpha_s(\mu^2)} \right]^{\delta \gamma_n^{(6)}} \left( {\mu^2 
       \over Q^2} \right)^2
       \label{eq:M1n}
 \ee
where the leading twist term $\delta\mu_n^{(1)}(Q^2)$ is given by Eqs. (\ref{eq:mu_large_n}-\ref{eq:a0}), while the logarithmic $pQCD$ evolution of the twist-4 (twist-6) contribution is accounted for by the term $[\alpha_s(Q^2) / \alpha_s(\mu^2)]^{\delta \gamma_1^{(4)}}$ ($[\alpha_s(Q^2) / \alpha_s(\mu^2)]^{\delta \gamma_1^{(6)}}$) with an effective anomalous dimension $\delta \gamma_1^{(4)}$ ($\delta \gamma_1^{(6)}$) and the parameter $\delta a_1^{(4)}$ ($\delta a_1^{(6)}$) represents the overall
strength of the twist-4 (twist-6) term at the renormalization scale $\mu^2$. For the latter we consider hereafter the value $\mu = 1 ~ GeV/c$ and, for fixing the running of the coupling constant $\alpha_s(Q^2)$, the updated $PDG$ value $\alpha_s(M_Z^2) = 0.118$ \cite{PDG} is adopted throughout this work.

\indent In Eq. (\ref{eq:M1n}) only twist-4 and twist-6 terms are included. In this respect we want to point out that the number of higher-twist terms to be considered is mainly governed by the $Q^2$-range of the analysis. Indeed, as the latter is extended down to lower and lower values of $Q^2$, more and more higher-twist terms are expected to contribute equally well. We anticipate here that for $Q^2 \gsim 1 ~ (GeV/c)^2$: ~ i) the inclusion of a twist-4 and a twist-6 term appears to work pretty well, as already found in case of the unpolarized moments \cite{Ricco,SIM00}, and ~ ii) our least-$\chi^2$ fitting procedure turns out to be not sufficiently sensitive for a precise determination of power corrections of order higher than the twist-6. Thus, just because of phenomenological findings we limit ourselves to consider only twist-4 and twist-6 terms in our analyses for $Q^2 \gsim 1 ~ (GeV/c)^2$.

\indent Let us start with the twist analysis of the first Nachtmann moment $M_1^{(1)}(Q^2)$; from Eqs. (\ref{eq:M1n}) and (\ref{eq:mu_first}) one has
 \be
       M_1^{(1)}(Q^2) & = & {<e^2> \over 2} \left[ \Delta q^{NS} + a_0(Q^2) 
       \right] \left[ 1 - {\alpha_s(Q^2) \over \pi} \right]  
       \nonumber \\
       & + & \delta a_1^{(4)} ~ \left[ {\alpha_s(Q^2) \over \alpha_s(\mu^2)} 
       \right]^{\delta \gamma_1^{(4)}} {\mu^2 \over Q^2} +  \delta 
       a_1^{(6)} ~ \left[ {\alpha_s(Q^2) \over \alpha_s(\mu^2)} 
       \right]^{\delta \gamma_1^{(6)}} \left( {\mu^2 \over Q^2} \right)^2.
       \label{eq:M1_first}
 \ee
The non-singlet moment $\Delta q^{NS}$ is taken fixed at the value $\Delta q^{NS} = 1.095$, deduced from the experimental values of the triplet and octet axial coupling constants [see Eq. (\ref{eq:Delta_NS})], with the latter obtained under the assumption of $SU(3)$-flavor symmetry. The unknown parameters in Eq. (\ref{eq:M1_first}) are the singlet axial charge $a_0(\mu^2)$ at the renormalization scale (or at any given value of $Q^2$) and the four higher-twist quantities $\delta a_1^{(4)}$, $\delta \gamma_1^{(4)}$, $\delta a_1^{(6)}$ and $\delta \gamma_1^{(6)}$. Their values, reported in Table 4, have been determined by fitting the pseudo-data of Fig. 7(a), adopting the least-$\chi^2$ procedure in the $Q^2$-range between $0.5$ and $50 ~ (GeV/c)^2$. The twist decomposition of the $Q^2$ behavior of the first moment $M_1^{(1)}(Q^2)$ is illustrated in Fig. 9.

\begin{table}[htb]

{\small {\bf Table 4.} Values of the parameters appearing in Eq. (\ref{eq:M1_first}) obtained by a least-$\chi^2$ procedure in the $Q^2$-range from $0.5$ to $50 ~ (GeV/c)^2$. The non-singlet moment is fixed at the value $\Delta q^{NS} = 1.095$ (see text). The last row reports the minimal value obtained for the $\chi^2$ variable divided by the number of degrees of freedom. The errors on the parameters represent the uncertainty of the fitting procedure corresponding to one-unit increment of the $\chi^2 / N_{d.o.f.}$ variable.}

\begin{center}

\begin{tabular}{||c||c|c||c|c||c||}
\hline \hline
 $a_0(10 ~ GeV^2)$ & $\delta a_1^{(4)}$ & $\delta \gamma_1^{(4)}$ & $\delta 
 a_1^{(6)}$ & $\delta \gamma_1^{(6)}$ & $\chi^2 / N_{d.o.f.}$
\\ \hline \hline
 $0.14 \pm 0.09$ & $0.038 \pm 0.012$ & $2.2 \pm 0.4$ & $-0.017 \pm 0.006$ & 
 $1.9 \pm 0.6$ & $0.053$
\\ \hline \hline
\end{tabular}

\end{center}

\end{table}

\begin{figure}[htb]

\centerline{\epsfxsize=14cm \epsfig{file=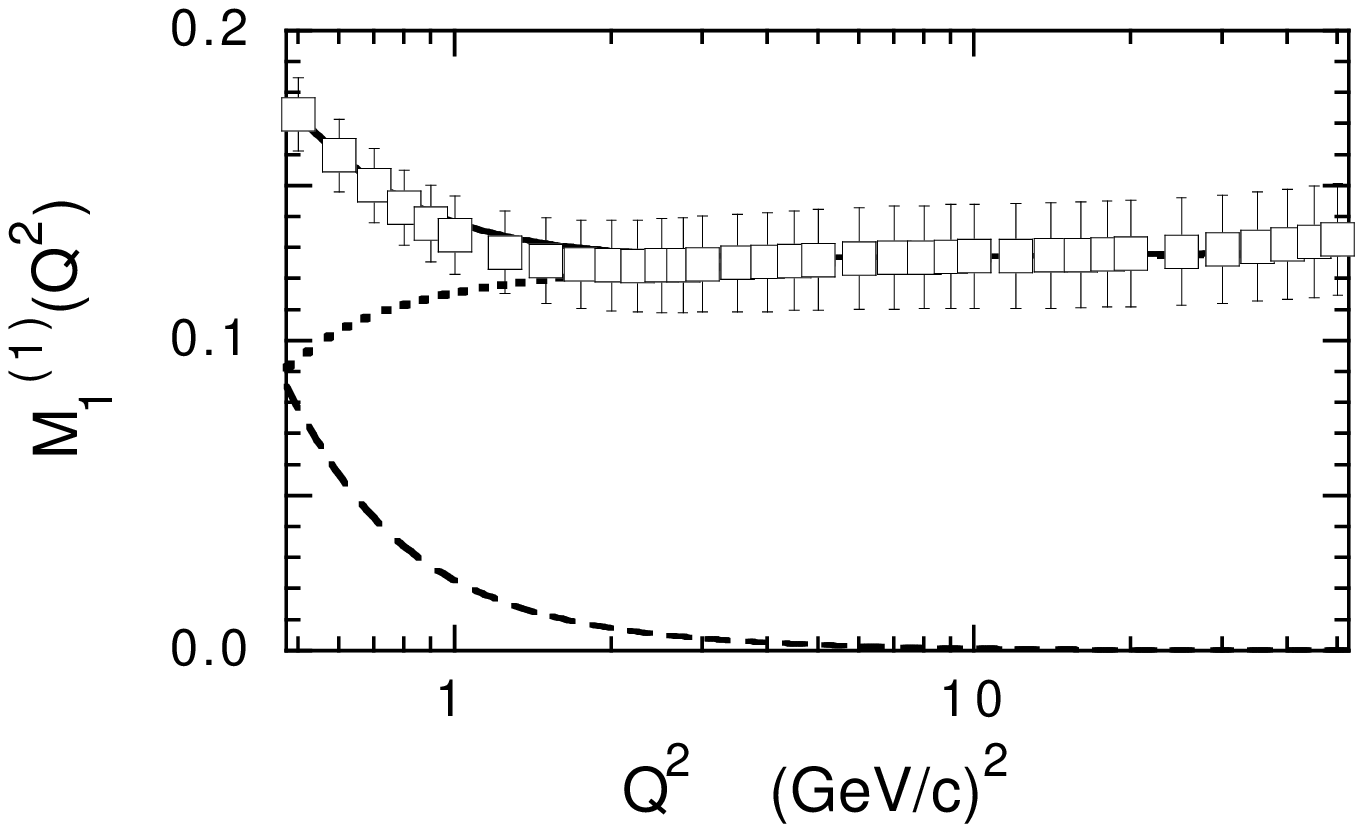}}

\small{ {\bf Figure 9}. Twist analysis of the first proton Nachtmann moment $M_1^{(1)}(Q^2)$. The solid line is the result of Eq. (\ref{eq:M1_first}) fitted by the least-$\chi^2$ procedure to our pseudo-data of Fig. 7(a) (open squares) in the $Q^2$-range between $0.5$ and $50 ~ (GeV/c)^2$. The dotted and dashed lines correspond to the contributions of the leading and total higher twists, given by the sum of the twist-4 and twist-6 terms in Eq. (\ref{eq:M1_first}), respectively.}

\end{figure}

\noindent From Table 4 and Fig. 9 it turns out that:

\begin{itemize}

\item{the total contribution of the higher twists is tiny for $Q^2 \gsim 1 ~ (GeV/c)^2$, but it is comparable with the leading twist already at $Q^2 \simeq 0.5 ~ (GeV/c)^2$. Since the first moment basically corresponds to the  area under the structure function $g_1^p$ (as it is the case of the second moment of the unpolarized structure function $F_2^p$), the dominance of the leading twist in $M_1^{(1)}(Q^2)$, occurring for $Q^2 \gsim 1 ~ (GeV/c)^2$, reflects only the concept of global duality and {\em not} that of local duality (cf. Ref. \cite{duality});}

\item{in our analysis, where the leading and the higher twists are simultaneously extracted, the singlet axial charge (in the $AB$ scheme)  is determined to be $a_0(10 ~ GeV^2) = 0.14 \pm 0.09$ (see Table 4), which nicely agrees with many recent estimates appeared in the literature, like e.g. $a_0(10 ~ GeV^2) = 0.10_{-0.11}^{+0.17}$ from Ref. \cite{Altarelli} and $a_0(10 ~ GeV^2) = 0.24 \pm 0.07 \mbox{(stat)} \pm 0.19 \mbox{(syst)}$ from Ref. \cite{SMC}(b). As a consistency check, we have limited our analysis to the high $Q^2$-range from $10$ to $50 ~ (GeV/c)^2$ including only the twist-2 contribution in Eq. (\ref{eq:M1_first}), obtaining $a_0(10 ~ GeV^2) = 0.18 \pm 0.09$. Thus, possible higher-twist effects on the extraction of $a_0$ are small and well within the uncertainties of the fitting procedure. We quote $a_0(10 ~ GeV^2) = 0.16 \pm 0.09$ as our final determination of the singlet axial charge in the $AB$ scheme. Our value of $a_0$ is therefore significantly below the naive quark-model expectation (i.e. compatible with the well known "proton spin crisis"), but it does not exclude completely a singlet axial charge as large as $\simeq 0.25$.}

\end{itemize}

\indent As explained in Section 2 [see Eq. (\ref{eq:mu_large_n})], for higher-order moments ($n \geq 3$) we make use of the following twist expansion
 \be
       M_{n \geq 3}^{(1)}(Q^2) & = & \delta A_n^(\mu^2) \left( 
       {\alpha_s(Q^2) \over \alpha_s(\mu^2)} \right)^{\gamma_n^{NS}} \left[ 
       1 + {\alpha_s(Q^2) \over 2 \pi} \delta C_n^{(q)} \right] \nonumber \\
       & \cdot & \left[ 1 + {\alpha_s(Q^2) - \alpha_s(\mu^2) \over 4 \pi} 
       \left( \gamma_n^{1, NS} - {\beta_1 \over \beta_0} \gamma_n^{NS} 
       \right) \right] \nonumber \\
       & + & \delta a_n^{(4)} ~ \left[ {\alpha_s(Q^2) \over \alpha_s(\mu^2)} 
       \right]^{\delta \gamma_n^{(4)}} {\mu^2 \over Q^2} +  \delta a_n^{(6)} 
       ~ \left[ {\alpha_s(Q^2) \over \alpha_s(\mu^2)} \right]^{\delta 
       \gamma_n^{(6)}} \left( {\mu^2 \over Q^2} \right)^2
       \label{eq:M1_large_n}
 \ee
The five parameters $\delta A_n^(\mu^2)$, $\delta a_n^{(4)}$, $\delta \gamma_n^{(4)}$, $\delta a_n^{(6)}$ and $\delta \gamma_n^{(6)}$ are simultaneously determined by the least-$\chi^2$ procedure applied to the pseudo-data of Fig. 7 in the $Q^2$-range from $1$ to $50 ~ (GeV/c)^2$; the obtained values are reported in Table 5, while the twist decomposition of $M_{n \geq 3}^{(1)}(Q^2)$ is illustrated in Fig. 10.

\begin{table}[htb]

{\small {\bf Table 5.} Values of the parameters appearing in Eq. (\ref{eq:M1_large_n}) at $\mu = 1 ~ (GeV/c)$, obtained by a least-$\chi^2$ procedure in the $Q^2$-range from $1$ to $50 ~ (GeV/c)^2$. The errors on the parameters represent the uncertainty of the fitting procedure corresponding to one-unit increment of the $\chi^2 / N_{d.o.f.}$ variable.}

\begin{center}

\begin{tabular}{||c||c|c|c|c||}
\hline \hline
 $parm.$ & $n = 3$ & $ n = 5$ & $n = 7$ & $n = 9$
\\ \hline \hline
 $\delta A_n$ & $0.133 \pm 0.019$ & $0.0297 \pm 0.0044$ & $0.0098 \pm 
 0.0013$ & $0.00343 \pm 0.00037$
\\ \hline \hline
 $\delta a_n^{(4)}$ & $0.015 \pm 0.005$ & $0.018 \pm 0.005$ & $0.026 \pm 
 0.003$ & $0.032 \pm 0.004$
\\ \hline
 $\delta \gamma_n^{(4)}$ & $1.7 \pm 0.6$ & $2.4 \pm 0.9$ & $3.7 \pm 0.5$ & 
 $4.6 \pm 0.5$
\\ \hline \hline
 $\delta a_n^{(6)}$ & $-0.006 \pm 0.002$ & $-0.014 \pm 0.005$ & $-0.024 \pm 
 0.004$ & $-0.031 \pm 0.005$
\\ \hline
 $\delta \gamma_n^{(6)}$ & $1.7 \pm 0.6$ & $1.7 \pm 0.5$ & $2.6 \pm 0.5$ & 
 $3.3 \pm 0.5$
\\ \hline \hline
 $\chi^2 / N_{d.o.f.}$ & $0.14$ & $0.39$ & $0.60$ & $0.82$
\\ \hline \hline
\end{tabular}

\end{center}

\end{table}

\begin{figure}[htb]

\centerline{\epsfxsize=16cm \epsfig{file=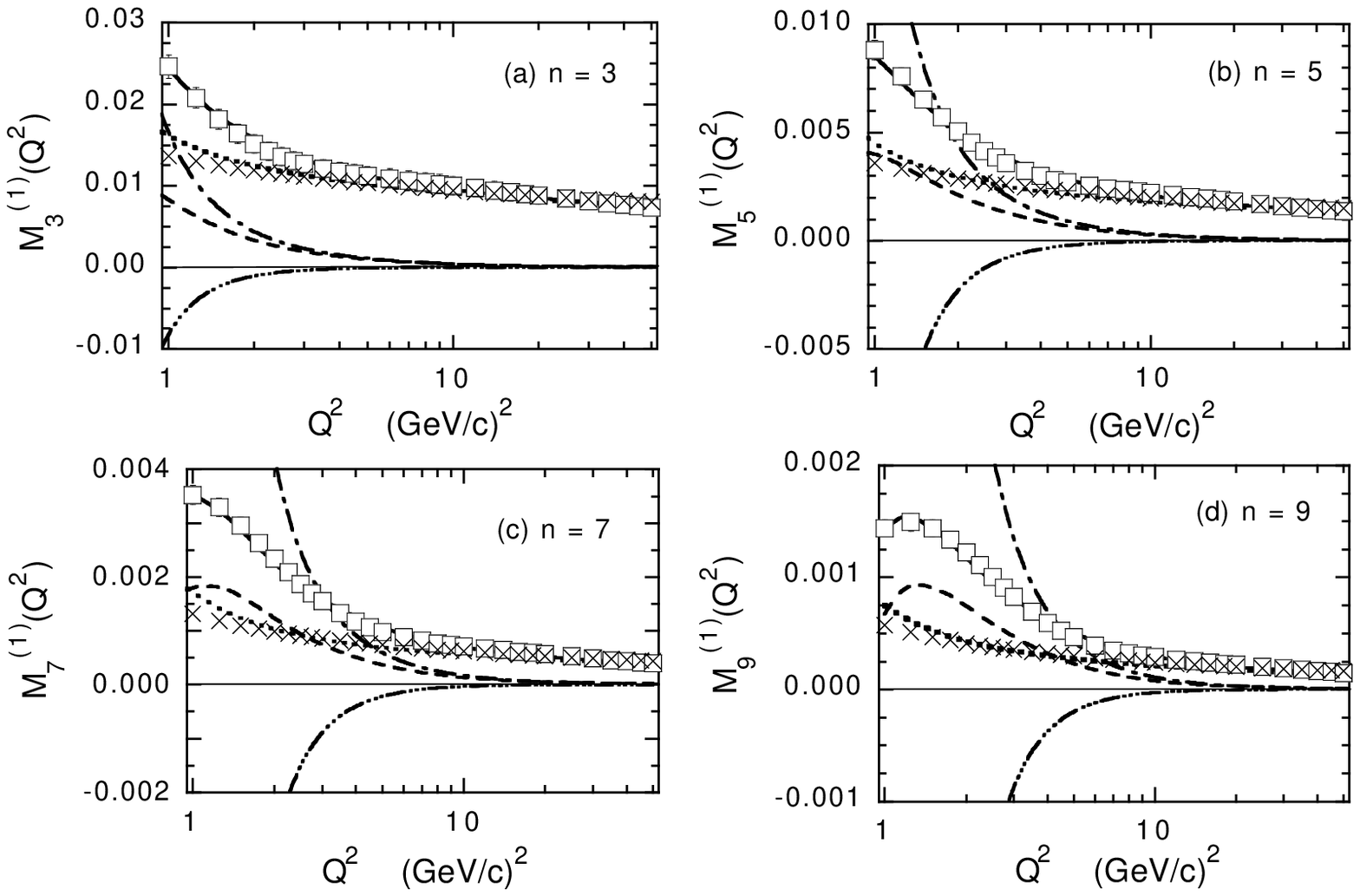}}

\small{ {\bf Figure 10}. Twist analysis of the proton Nachtmann moments $M_n^{(1)}(Q^2)$ with $n = 3$ (a), $n = 5$ (b), $n = 7$ (c) and $n = 9$ (d). The solid line is the result of Eq. (\ref{eq:M1_large_n}) fitted by the least-$\chi^2$ procedure to our pseudo-data (open squares). The dotted, dot-dashed and triple-dot-dashed lines are the separate contribution in Eq. (\ref{eq:M1_large_n}) of the twist-2, twist-4 and twist-6 terms, respectively. The dashed line is the total higher-twist contribution, given by sum of the twist-4 and twist-6 terms. The crosses are the moments calculated using the $NLO$ polarized $PDF$ set of Ref. \cite{GRSV}, labeled valence scenario.}

\end{figure}

\indent Our main results can be summarized as follows:

\begin{itemize}

\item the twist-2 term, extracted from our proton pseudo-data simultaneously with the twist-4 and twist-6 contributions, differs only slightly from the predictions obtained using the polarized $PDF$ set of Ref. \cite{GRSV} evolved at $NLO$ (compare crosses and dotted lines in Fig. 10). A similar situation holds as well in case of the $PDF$ set of Ref. \cite{GS96}. Moreover, we have checked that, by limiting our analysis to the $Q^2$-range from $10$ to $50 ~ (GeV/c)^2$ and without including any higher-twist term, our extracted twist-2  changes only within the fitting uncertainties reported in Table 5; 

\item the twist-4 and twist-6 contributions turn out to have opposite signs, making the total higher-twist contribution smaller than its individual terms, as we have already found in case of the unpolarized proton and deuteron structure functions (see Ref. \cite{Ricco}. As $n$ increases (i.e. as $x$ increases), the total higher-twist contribution increases and becomes comparable or even larger than the leading twist term for $Q^2 \sim$ few $(GeV/c)^2$ (compare dashed and dotted lines in Fig. 10);

\item for $n = 3 \div 9$ the values of the effective anomalous dimensions $\delta \gamma_n^{(4)}$ (and to a less extent also the values of $\delta \gamma_n^{(6)}$)  result to be larger than the values of the corresponding twist-2 anomalous dimensions ($\gamma_n^{NS} \simeq 0.6 \div 1.2$ for $n = 3 \div 9$ \cite{Gual});

\item the uncertainties on the different twist contributions due to our least-$\chi^2$ procedure are within $\simeq 40 \%$ for the twist-4 and twist-6 terms, while they are about $10 \div 15 \%$ for the leading twist (see Table 5);

\item the twist expansion (\ref{eq:M1_large_n}) appears to work quite well for values of $Q^2$ down to $\simeq 1 ~ (GeV/c)^2$.

\end{itemize}

\section{Soft gluon resummation and higher twists}

\indent So far the power corrections appearing in Eqs. (\ref{eq:M1_first}-\ref{eq:M1_large_n}) have been extracted from our pseudo-data assuming the $NLO$ approximation for the leading twist, and therefore they represent the higher twists at $NLO$. Since our main aim is to get information on the dynamical power corrections generated by multiparton correlations, it is necessary to estimate the possible effects of higher orders of the perturbative series, which defines the twist-2 coefficient functions $E_{n2}[\mu, \alpha_s(Q^2)]$ appearing in Eq. (\ref{eq:CN}). Such a job has been already carried out in case of the unpolarized proton structure function  $F_2^p(x, Q^2)$ in Refs. \cite{Ricco} and \cite{SIM00}. In the former the high-order perturbative terms have been estimated through the ambiguities introduced by infrared renormalons, while in the latter the effects of soft gluon emission have been taken into account at large $x$ via well-established soft-gluon resummation techniques \cite{SGR}. In both cases the comparison of the results obtained adopting the $NLO$ approximation and those including high-order corrections has clearly shown that the extraction of higher twists at large $x$ is remarkably sensitive to soft gluon effects. Therefore, in this Section we address the issue of high-order effects by extending the calculations of Ref. \cite{SIM00} to the Nachtmann moments of the polarized proton structure function $g_1^p(x, Q^2)$.

\indent Let us start from Eq. (\ref{eq:M1_large_n}) [see also Eq. (\ref{eq:mu_large_n})] where $\delta C_n^{(q)}$ is the $NLO$ part of the quark coefficient function. In the $\overline{MS}$ (or $AB$) scheme it reads explicitly as
 \be
       \delta C_n^{(q)} = C_F \left\{ S_1(n) \left[ S_1(n) + {3 \over 2} - 
       {1 \over n(n+1)} \right] - S_2(n) + {1 \over 2n} + {1 \over n+1} + {1 
       \over n^2} - {9 \over 2} \right\}
       \label{eq:delta_Cn_NLO}
 \ee
where $C_F \equiv (N_c^2 - 1) / (2 N_c) \to 4 / 3$ and $S_k(n) \equiv \sum_{j=1}^n 1 /j^k$. Note that $\delta C_n^{(q)}$ differs from the corresponding quark coefficient function of the unpolarized case (cf., e.g.,  Eq. (4) of Ref. \cite{SIM00}) only in the terms proportional to $1 / n$ and $1 / (n + 1)$ in the r.h.s. of Eq. (\ref{eq:delta_Cn_NLO}). Thus, the $SGR$ effects can be included in the polarized $NS$ moments in a way completely similar to the unpolarized $NS$ case already carried out in Ref. \cite{SIM00}. For completeness here below we report the explicit expressions of $SGR$ effects in the polarized $NS$ case.

\indent For large $n$ (corresponding to the large-$x$ region) the coefficient $\delta C_n^{(q)}$ is logarithmically divergent. Indeed, since $S_1(n) = \gamma_E + \mbox{log}(n) + O(1 / n)$ and $S_2(n) = \pi^2 / 6 + O(1 / n)$ (with $\gamma_E$ being the Euler-Mascheroni constant), one gets
 \be
       \delta C_n^{(q)} = \delta C_{DIS}^{(q)} + \delta C_{n, LOG}^{(q)} + 
       O(1 / n) 
       \label{eq:delta_Cn_exp}
 \ee
where
 \be
       \label{eq:delta_Cn_DIS}
       \delta C_{DIS}^{(q)} & = & C_F \left[ \gamma_E^2 + {3 \over 2} 
       \gamma_E - {9 \over 2} - {\pi^2 \over 6} \right] ~ , \\
       \delta C_{n, LOG}^{(q)} & = & C_F ~ \mbox{ln}(n) \left[ \mbox{ln}(n) 
       + 2\gamma_E + {3 \over 2} \right]
       \label{eq:delta_Cn_LOG}
 \ee

\indent The physical origin of the logarithm and double logarithm terms in Eq. (\ref{eq:delta_Cn_LOG}) is the mismatch among the singularities generated in the quark coefficient function by the virtual gluon loops and the real gluon emissions, the latter being suppressed as the elastic peak corresponding to the threshold $x = 1$ is approached. In other words at large $x$ the relevant scale is no more $Q^2$, but $Q^2 (1 - x)$ \cite{Brodsky} and the usual Altarelli-Parisi evolution equation should be accordingly modified \cite{Amati}. The presence of the abovementioned divergent terms at large $n$ would spoil the perturbative nature of the $NLO$ approximation (as well as of any fixed-order calculation) and therefore the effects of soft gluon emissions should be considered at all orders in the strong coupling constant $\alpha_s$. To this end one can take advantage of resummation techniques, which show that in moment space soft gluon effects exponentiate \cite{SGR,Catani,Vogt}. Thus, the moments of the leading twist including $SGR$ effects acquire the following form
 \be
       \delta \mu_{n \geq 3}^{(1, SGR)}(Q^2) & = & \delta 
       \overline{A}_n(\mu^2) \left[ {\alpha_s(Q^2) \over \alpha_s(\mu^2)} 
       \right]^{\gamma_n^{NS}}
       \nonumber \\
       & \cdot & \left\{ \left[ 1 + {\alpha_s(Q^2) \over 2 \pi} 
       C_{DIS}^{(q)} \right]  e^{G_n(Q^2)} + {\alpha_s(Q^2) \over 2 \pi} 
       \left[ C_n^{(q)} -  C_{DIS}^{(q)} - C_{n, LOG}^{(q)} \right]  
       \right\} \nonumber \\
       & \cdot & \left\{ 1 + {\alpha_s(Q^2) - \alpha_s(\mu^2)  \over 4 \pi} 
       \left( \gamma_n^{1, NS} - {\beta_1 \over \beta_0} \gamma_n^{NS} 
       \right) \right\}
       \label{eq:mu_SGR}
 \ee
where the function $G_n(Q^2)$ is the key quantity of the soft gluon resummation and reads as (cf. \cite{Catani,Vogt})
 \be
       G_n(Q^2) = \int_0^1 dz {z^{n-1} - 1 \over 1 - z} \left\{ {1 \over 2} 
       B[\alpha_s(Q^2(1 - z))] + \int_{Q^2}^{Q^2(1 - z)} {dq^2 \over q^2} 
      A[\alpha_s(q^2)] \right\}
      \label{eq:Gn_def}
 \ee
where $A[\alpha_s] = C_F \alpha_s / \pi + C_F K (\alpha_s / \pi)^2 / 2$, $B[\alpha_s] = - 3 C_F \alpha_s / 2\pi$ with $K = C_A (67/18 - \pi^2 / 6) - 10 T_R N_f / 9$, $C_A = N_c \to 3$ and $T_R = 1/2$. Explicitly one has
 \be
       G_n(Q^2) = \mbox{ln}(n) ~ G_1(\lambda_n) + G_2(\lambda_n) + 
       O[\alpha_s^k \mbox{ln}^{k-1}(n)]
       \label{eq:Gn}
 \ee
where $\lambda_n \equiv \beta_0 \alpha_s(Q^2) \mbox{ln}(n) / 4\pi$ and
 \be
       G_1(\lambda) & = & C_F {4 \over \beta_0 \lambda} \left[ \lambda + (1 
       -  \lambda) \mbox{ln}(1 - \lambda) \right] ~ , \nonumber \\
       G_2(\lambda) & = & - C_F {4 \gamma_E + 3 \over \beta_0} \mbox{ln}(1 
       - \lambda) - C_F {8 K \over \beta_0^2} \left[ \lambda + \mbox{ln}(1 
       - \lambda) \right] + \nonumber \\
       & & C_F {4  \beta_1 \over \beta_0^3} \left[ \lambda + \mbox{ln}(1 - 
       \lambda) + {1 \over 2} \mbox{ln}^2(1 - \lambda) \right]
      \label{eq:G1G2}
 \ee
It is straightforward to check that in the limit $\lambda_n << 1$ one has $G_n(Q^2) \to \alpha_s(Q^2) C_{n, LOG}^{(q)} / 2\pi$, so that at $NLO$ Eq. (\ref{eq:mu_SGR}) reduces to Eq. (\ref{eq:mu_large_n}). Note that the function $G_2(\lambda)$ is divergent for $\lambda \to 1$; this means that at large $n$ (i.e. large $x$) the soft gluon resummation cannot be extended to arbitrarily low values of $Q^2$. Therefore, for a safe use of present $SGR$ techniques we will work far from the above-mentioned divergencies by limiting our analyses of low-order moments ($n \leq 9$) to $Q^2 \geq 1 ~ (GeV/c)^2$ \footnote{We want to stress that, as in Refs. \cite{Ricco,SIM00}, our main aim is not to perform a full calculation of perturbative corrections, but to check whether the $NLO$ approximation can provide a safe extraction of higher twists.}. 

 \indent The results obtained for the ratio of the quark coefficient function calculated within the $SGR$ technique and at $NLO$, namely
 \be
       r_n(Q^2) = {\left( 1 + \alpha_s(Q^2) C_{DIS}^{(q)} / 2\pi \right) 
       e^{G_n(Q^2)} + \alpha_s(Q^2) \left( C_n^{(q)} - C_{DIS}^{(q)} - C_{n, 
       LOG}^{(q)} \right) / 2\pi \over 1 + \alpha_s(Q^2) ~ C_n^{(q)} / 
       2\pi},
      \label{eq:ratio}
 \ee
are reported in Fig. 11 at $\alpha_s(M_Z^2) = 0.118$. It can be seen that soft gluon effects are quite small for the first and the third moments, while for $n \geq 5$ they increase significantly as $n$ increases, particularly for $Q^2 \sim$ few $(GeV/c)^2$. Note that for $n = 1$ one has $r_1(Q^2) = 1$ at any values of $Q^2$ and $\alpha_s(M_Z^2)$; therefore, no effect from soft gluons can occur in the first moment $M_1^{(1)}(Q^2)$.

\begin{figure}[htb]

\centerline{\epsfxsize=14cm \epsfig{file=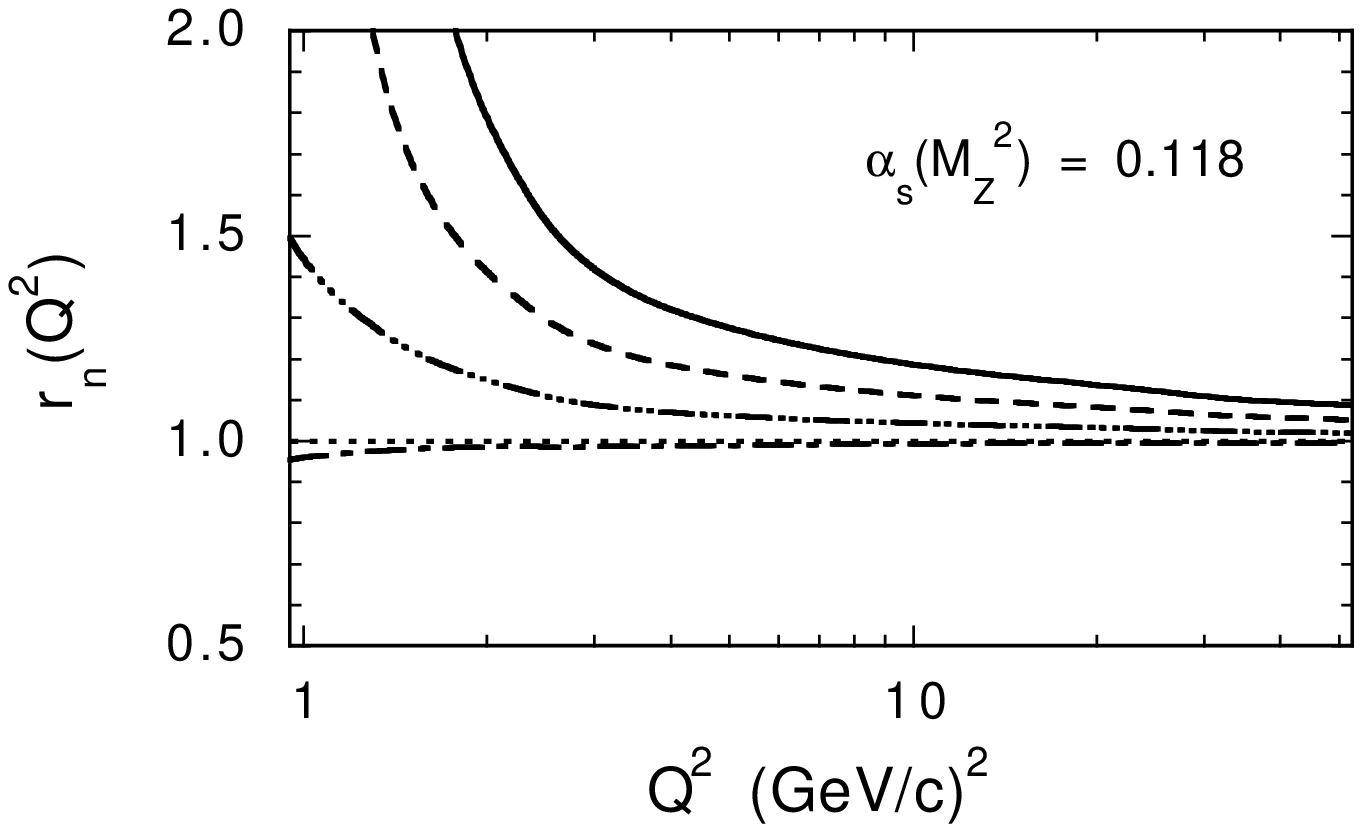}}

\small{ {\bf Figure 11}. Values of the ratio $r_n(Q^2)$ [see Eq. (\ref{eq:ratio})] of the quark coefficient function within the $SGR$ technique and at $NLO$. The dotted, dot-dashed, triple-dot-dashed, dashed and solid lines correspond to $n = 1, 3, 5, 7$ and $9$, respectively. Note that $r_1(Q^2) = 1$ and $r_3(Q^2) < 1$.}

\end{figure}

\indent Thus, for $n \geq 3$  our pseudo-data of Fig. 7 have been analyzed in the $Q^2$-range from $1$ to $50 ~ (GeV/c)^2$ including soft-gluon effects, viz
 \be
       M_n^{(1)}(Q^2) = \delta \mu_n^{(1, SGR)}(Q^2) + \delta 
       \overline{a}_n^{(4)} ~ \left[ {\alpha_s(Q^2) \over \alpha_s(\mu^2)} 
       \right]^{\delta \overline{\gamma}_n^{(4)}} {\mu^2 \over Q^2} + \delta 
       \overline{a}_n^{(6)} ~ \left[ {\alpha_s(Q^2) \over \alpha_s(\mu^2)} 
       \right]^{\delta \overline{\gamma}_n^{(6)}} \left( {\mu^2 \over Q^2} 
       \right)^2
       \label{eq:M1n_SGR}
 \ee
where the leading twist term $\delta \mu_n^{(1, SGR)}(Q^2)$ is given by Eq. (\ref{eq:mu_SGR}). The five parameters $\delta \overline{A}_n(\mu^2)$, $\delta \overline{a}_n^{(4)}$, $\delta \overline{\gamma}_n^{(4)}$, $\delta \overline{a}_n^{(6)}$ and $\delta \overline{\gamma}_n^{(6)}$ are simultaneously determined by the least-$\chi^2$ procedure; the obtained values are reported in Table 6, while the twist decomposition of $M_{n \geq 3}^{(1)}(Q^2)$ is illustrated in Fig. 12. It can be seen that the twist-2 parameters $\delta \overline{A}_n$ (see Table 6) almost coincide with the corresponding $NLO$ quantities $\delta A_n$ (see Table 5) within the uncertainties generated by our least-$\chi^2$ procedure. As in the case of the $NLO$ analysis, the twist-4 and twist-6 contributions are again well defined and with opposite signs, making the total higher-twist contribution smaller than its individual terms. The strengths $\delta \overline{a}_n^{(4)}$ and $\delta \overline{a}_n^{(6)}$ (see Table 6) differ remarkably from the corresponding $NLO$ quantities $\delta a_n^{(4)}$ and $\delta a_n^{(6)}$ (see Table 5), while the effective anomalous dimensions $\delta \overline{\gamma}_n^{(4)}$ and $\delta \overline{\gamma}_n^{(6)}$ do not change significantly. We point out that at large $n$  (i.e. at large $x$), the total higher-twist contribution is reduced by soft gluon effects, but it is still a significant fraction of the leading twist term for $Q^2 \sim$ few $(GeV/c)^2$ (compare dashed and dotted lines in Fig. 12).

\begin{table}[htb]

{\small {\bf Table 6.} Values of the parameters appearing in Eq. (\ref{eq:M1n_SGR}) at $\mu = 1 ~ (GeV/c)$, obtained by a least-$\chi^2$ procedure in the $Q^2$-range from $1$ to $50 ~ (GeV/c)^2$. The errors are as in Table 5.}

\begin{center}

\begin{tabular}{||c||c|c|c|c||}
\hline \hline
 $parm.$ & $n = 3$ & $ n = 5$ & $n = 7$ & $n = 9$
\\ \hline \hline
 $\delta \overline{A}_n$ & $0.135 \pm 0.025$ & $0.0297 \pm 0.0050$ & $0.0092 
 \pm 0.0016$ & $0.00358 \pm 0.00064$
\\ \hline \hline
 $\delta \overline{a}_n^{(4)}$ & $0.015 \pm 0.005$ & $0.0080 \pm 0.0025$ & 
 $0.0067 \pm 0.0008$ & $0.0080 \pm 0.0009$
\\ \hline
 $\delta \overline{\gamma}_n^{(4)}$ & $1.9 \pm 0.7$ & $1.9 \pm 0.6$ & $2.4 
 \pm 0.9$ & $4.0 \pm 1.5$
\\ \hline \hline
 $\delta \overline{a}_n^{(6)}$ & $-0.0062 \pm 0.0022$ & $-0.0053 \pm 0.0019$ 
 & $-0.0073 \pm 0.0007$ & $-0.0011 \pm 0.0002$
\\ \hline
 $\delta \overline{\gamma}_n^{(6)}$ & $1.9 \pm 0.8$ & $3.0 \pm 1.1$ & $3.5 
 \pm 1.2$ & $3.7 \pm 1.3$
\\ \hline \hline
 $\chi^2 / N_{d.o.f.}$ & $0.13$ & $0.29$ & $0.46$ & $0.67$
\\ \hline \hline
\end{tabular}

\end{center}

\end{table}

\begin{figure}[htb]

\centerline{\epsfxsize=16cm \epsfig{file=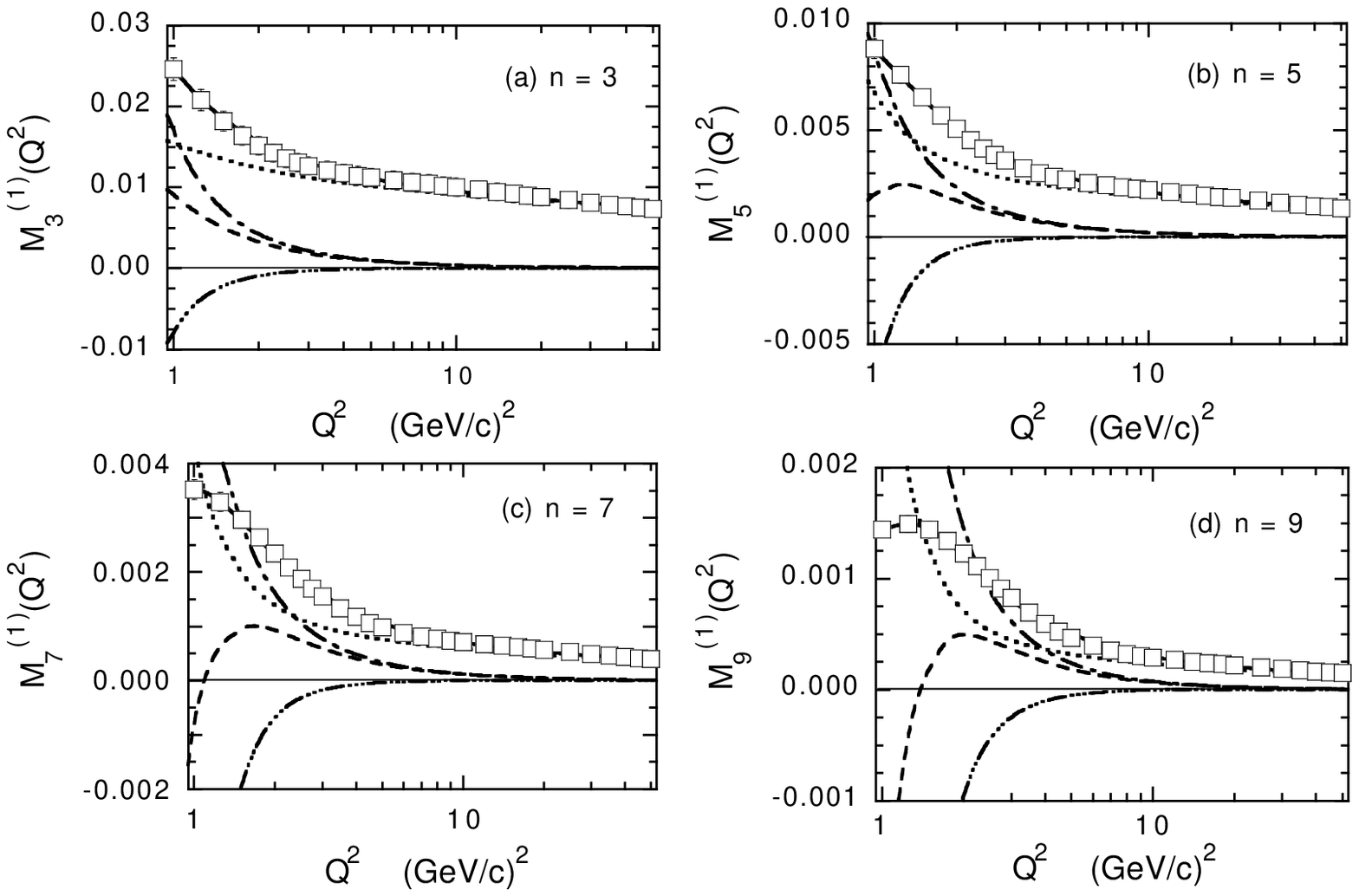}}

\small{ {\bf Figure 12}. Twist analysis of the proton Nachtmann moments $M_n^{(1)}(Q^2)$ with $n = 3$ (a), $n = 5$ (b), $n = 7$ (c) and $n = 9$ (d). The solid line is the result of Eq. (\ref{eq:M1n_SGR}) fitted by the least-$\chi^2$ procedure to our pseudo-data (open squares). The dotted, dot-dashed and triple-dot-dashed lines are the separate contribution in Eq. (\ref{eq:M1n_SGR}) of the twist-2, twist-4 and twist-6 terms, respectively. The dashed line is the total higher-twist contribution, given by sum of the twist-4 and twist-6 terms.}

\end{figure}

\indent The comparison of our twist analyses at $NLO$ and within the $SGR$ technique is reported in Fig. 13, where it can be seen that, except for the third moment (which is the only one characterized by $r_3(Q^2) < 1$), the contribution of the twist-2 is enhanced by soft gluon effects, while the total higher-twist term decreases significantly after the resummation of soft gluons. Thus, as already observed  in the unpolarized case in Ref. \cite{SIM00}, also in the polarized one it is mandatory to go beyond the $NLO$ approximation and to include soft gluon effects in order to achieve a safer extraction of higher twists at large $x$, particularly for $Q^2 \sim$ few $(GeV/c)^2$.

\begin{figure}[htb]

\centerline{\epsfxsize=16cm \epsfig{file=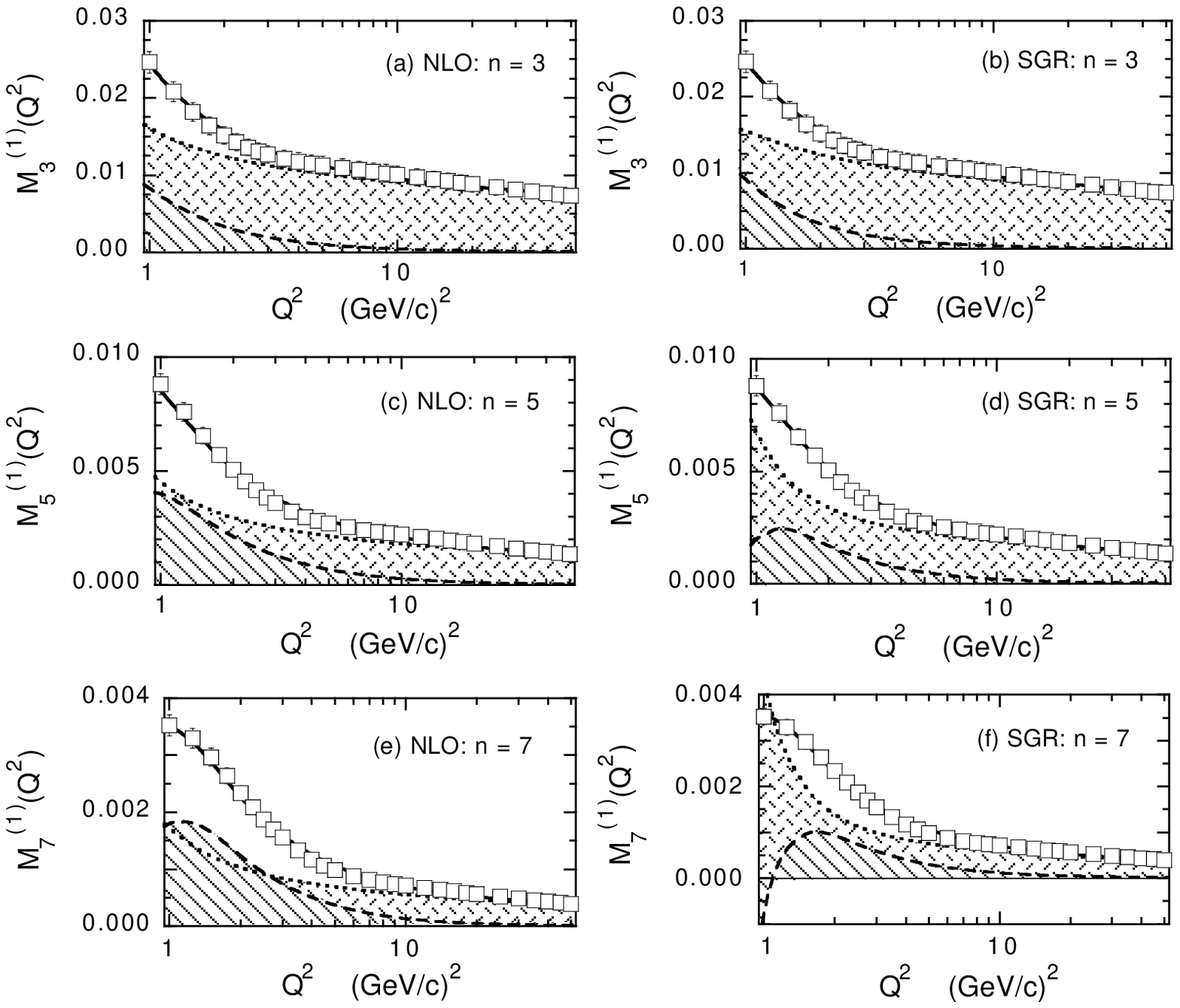}}

\small{ {\bf Figure 13}. Comparison of the twist analyses of the proton polarized Nachtmann moments $M_{n \geq 3}^{(1)}(Q^2)$ at $NLO$ [see Eq. (\ref{eq:M1_large_n})] and within the $SGR$ technique [see Eq. (\ref{eq:M1n_SGR})] for $n = 3$ (a,b), $n = 5$ (c,d) and $n = 7$ (e,f). Dotted lines are the twist-2 term, while dashed lines correspond to the total higher-twist contribution, given by sum of the twist-4 and twist-6 terms. The shaded areas help to identify the separate contributions of the leading and higher twists.}

\end{figure}

\indent In Fig. 14 we have compared the twist decomposition of the polarized Nachtmann moments $M_n^{(1)}(Q^2)$ (with $n = 3, 5, 7$) obtained in this work and the corresponding decomposition of the unpolarized (transverse) Nachtmann moments $M_n^{(T)}(Q^2)$ (with $n = 4, 6, 8$) obtained in Ref. \cite{SIM00} adopting the same $SGR$ technique. It can clearly be seen that our extracted higher-twist contribution appears to be a larger fraction of the leading twist in case of the polarized moments. This findings suggests that spin-dependent multiparton correlations may have  more impact than spin-independent ones.

\begin{figure}[htb]

\centerline{\epsfxsize=16cm \epsfig{file=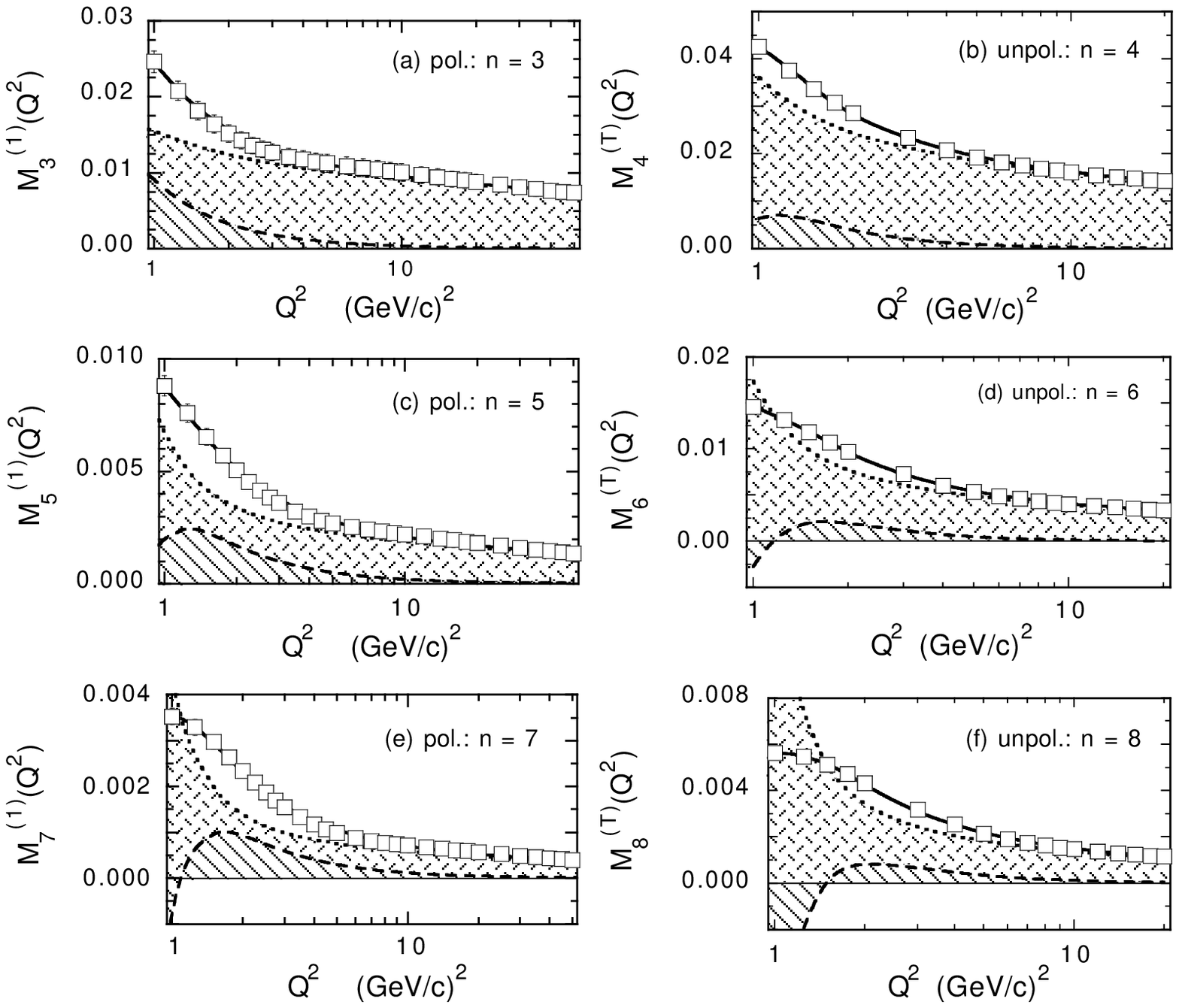}}

\small{ {\bf Figure 14}. Comparison of the twist analyses of the proton polarized $M_n^{(1)}(Q^2)$ (a,c,e) and unpolarized $M_n^{(T)}(Q^2)$ (b,d,f) Nachtmann moments adopting the $SGR$ technique resulting from this work and from Ref. \cite{SIM00}, respectively. The meaning of the lines and shaded areas is the same as in Fig. 13. Note that the scale of the vertical axis is different for polarized and unpolarized moments. }

\end{figure}

\section{Conclusions}

\indent In conclusion, we have extend the twist analysis made in Refs. \cite{Ricco} and \cite{SIM00} to the case of the polarized proton structure function $g_1^p(x, Q^2)$. Within the framework of the Operator Product Expansion we have used the Nachtmann moments in order to disentangle the kinematical target-mass corrections from the dynamical higher-twist effects related to correlations among partons. Since the evaluation of the Nachtmann moments requires the knowledge of the polarized structure functions in the whole $x$-range for fixed values of $Q^2$, we have developed a new parameterization of $g_1^p(x, Q^2)$, which describes the $DIS$ proton data up to $Q^2 \sim 50 ~ (GeV/c)^2$ and includes a phenomenological Breit-Wigner ansatz able to reproduce the existing electroproduction data in the proton-resonance regions.  Our interpolation formula for $g_1^p(x, Q^2)$ has been successfully extended down to the photon point, showing that it can nicely reproduce the very recent data \cite{Mainz} on the energy dependence of the asymmetry of the transverse photoproduction cross section as well as the experimental value of the proton Drell-Hearn-Gerasimov sum rule. According to our parameterization of $g_1^p(x, Q^2)$ the generalized Drell-Hearn-Gerasimov sum rule is predicted to have a zero-crossing point at $Q^2 = 0.16 \pm 0.04 ~ (GeV/c)^2$. 

\indent Low-order polarized Nachtmann moments have been evaluated and their $Q^2$ behavior has been investigated in terms of leading and higher twists for $Q^2 \gsim 1 ~ (GeV/c)^2$. In our analyses the leading twist is extracted simultaneously with a phenomenological higher-twist term from our pseudo-data. We have reported results obtained both at $NLO$ and adopting the same soft-gluon resummation technique applied to the analyses of the unpolarized data made in Ref. \cite{SIM00} in order to take into account the effects of higher orders of the perturbative series. As far as the first moment is concerned, the effects of higher twists are found to be quite small for $Q^2 \gsim 1 ~ (GeV/c)^2$. Moreover, the singlet axial charge is determined to be $a_0[10 ~ (GeV/c)^2] = 0.16 \pm 0.09$; our extracted value is significantly below the naive quark-model expectation (i.e., compatible with the well-known "proton spin crisis"), but it does not exclude completely a value of the singlet axial charge as large as $\simeq 0.25$. In case of higher order moments, which are more sensitive to the large-$x$ region, higher-twist effects are significantly reduced by the introduction of soft gluon contributions, but they are still relevant for $Q^2 \sim$ few $(GeV/c)^2$ at variance with the case of the unpolarized transverse structure function of the proton examined in Ref. \cite{SIM00}. Our finding suggests that spin-dependent correlations among partons may have more impact than spin-independent ones. As a byproduct, it has also been shown that the Bloom-Gilman local duality is strongly violated in the region of polarized electroproduction of the $\Delta(1232)$ resonance.

\indent The analysis of deuteron data, aimed at the determination of the flavor dependence of leading and higher twists, is in progress and the results will be presented elsewhere. 

\indent The results presented in this work and those already reported in Refs. \cite{Ricco} and \cite{SIM00} demonstrate that power correction terms can be safely extracted from a proper analysis of inclusive data. In particular, our extracted values of the effective strengths of the twist-4 and twist-6 terms may be directly compared with theoretical results obtained from first-principle calculations (like lattice $QCD$ simulations) or from models of the nucleon structure. In this way the non-perturbative regime of $QCD$ may be tested.

\indent Let us finally stress that our present analysis is mainly limited by the use of phenomenological fits of existing data (i.e., {\em pseudo-data}), which are required for interpolating smoothly the nucleon structure functions in the whole $x$-range for fixed values of $Q^2$. Therefore, polarized inclusive data with better quality at $x \gsim 0.5$ and $Q^2 \lsim 10 ~ (GeV/c)^2$, which may be collected at planned facilities like, e.g., $JLAB ~ @ ~ 12 ~ GeV$, could greatly help to improve our understanding of the non-perturbative structure of the nucleon. Finally, we want to point out that, since in inclusive polarized and unpolarized processes multiparton correlations appear to generate power-like terms with opposite signs, semi-inclusive or exclusive experiments might offer the possibility to achieve a better sensitivity to individual non-perturbative power corrections.

\section*{Acknowledgments} The authors gratefully acknowledge the $GDH$ and $A2$ collaborations for providing us the experimental results of Ref. \cite{Mainz}, as well as M. Stratmann and T. Gehrmann for supplying us with the fortran code for the evaluation of the polarized $PDF$ sets of Refs. \cite{GRSV,GS96}, respectively.


\begin{thebibliography}{99}

\bibitem{Ricco} G. Ricco, S. Simula and M. Battaglieri: Nucl. Phys. {\bf 
 B555} (1999) 306. G. Ricco and S. Simula: Nucl. Phys. {\bf A663\&664} 
 (2000) 1015; Nucl. Phys. {\bf A666\&667} (2000) 165; in Proc. of the Int'l 
 Workshop on {\em JLAB: Physics and Instrumentation with 6-12 GeV Beams}, 
 Jefferson Laboratory (Newport News, USA), June 1998, JLab Press (Newport 
 News, 1999), p. 313 (also e-print archive hep-ph/9809264).

\bibitem{Bodek} U.K. Yang and A. Bodek: Phys. Rev. Lett. {\bf 82} (1999) 
 2467.

\bibitem{Ji} X. Ji and P. Unrau: Phys. Rev. {\bf D52} (1995) 72.

\bibitem{Kataev} A.L. Kataev et al.: Phys. Lett. {\bf B388} (1996) 179; {\em 
 ib.} {\bf B417} (1998) 374; Nucl. Phys. {\bf A666\&667} (2000) 179. S.I. 
 Alekhin and A.L. Kataev: Nucl. Phys. {\bf A666\&667} (2000) 184.

\bibitem{renormalons} M. Dasgupta and B.R. Webber: Phys. Lett. {\bf B382} 
 (1996) 273. E. Stein et al.: Phys. Lett. {\bf B376} (1996) 177. M. Maul et 
 al.: Phys. Lett. {\bf B401} (1997) 100.

\bibitem{SIM00} S. Simula: Phys. Lett. {\bf B493} (2000) 325.

\bibitem{SGR} G. Sterman: Nucl. Phys. {\bf B281} (1987) 310. S. Catani and 
 L. Trentadue: Nucl. Phys. {\bf B327} (1989) 323.

\bibitem{PDG} Particle Data Group, D.E. Groom et al.: Eur. Phys. Jour. {\bf 
 C15} (2000) 1.

\bibitem{duality} For a proper updated analysis of local duality in the 
 unpolarized nucleon and nuclear structure functions see: G. Ricco et al.: 
 Phys. Rev. {\bf C57} (1998) 356; Few-Body Syst. Suppl. {\bf 10} (1999) 423. 
 S. Simula: Phys. Lett. {\bf B481} (2000) 14; e-print archive 
 hep-ph/0012193.

\bibitem{OPE} H.D. Politzer: Phys. Rev. Lett. {\bf 30} (1973) 1346. D.J. 
 Gross and F. Wilczek: Phys. Rev. Lett. {\bf 30} (1973) 1323.

\bibitem{Nachtmann} O. Nachtmann: Nucl. Phys. {\bf B63} (1973) 237.

\bibitem{Wandzura} S. Wandzura: Nucl. Phys. {\bf B122} (1977) 412.

\bibitem{Uematsu} S. Matsuda and T. Uematsu: Nucl. Phys. {\bf B168} (1980) 
181.

\bibitem{Mainz} J. Ahrens et al.: Phys. Rev. Lett. {\bf 87} (2001) 022003.

\bibitem{DHG} S.D. Drell and A.C. Hearn: Phys. Rev. Lett. {\bf 16} (1966) 
 908. S.B. Gerasimov: Sov. J. Nucl. Phys. {\bf 2} (1966) 430.

\bibitem{Altarelli} G. Altarelli et al.: Nucl. Phys. {\bf B496} (1997) 337.

\bibitem{BG} E. Bloom and F. Gilman: Phys. Rev. Lett. {\bf 25} (1970) 1140; 
 Phys. Rev. {\bf D4} (1971) 2901.

\bibitem{NLO} R. Mertig and W. van Neerven: Z. Phys. {\bf C70} (1996) 637. 
 W. Vogelsang: Phys. Rev. {\bf D54} (1996) 2023.

\bibitem{BFR96} R. Ball, S. Forte and G. Ridolfi: Phys. Lett. {\bf B378} 
 (1996) 255.

\bibitem{CR93} F.E. Close and R.G. Roberts: Phys. Lett. {\bf B316} (1993) 
 165.

\bibitem{E143} $E143$ collaboration, K. Abe et al.: Phys. Rev. {\bf D58} 
 (1998) 112003 and references therein quoted.

\bibitem{SMC} $SMC$ collaboration, (a) P. Adams et al.: Phys. Rev. {\bf D56} 
 (1997) 5330. (b) B. Adeva et al.: Phys. Rev. {\bf D58} (1998) 112001.

\bibitem{E130} $E130$ collaboration, G. Baum et al.: Phys. Rev. Lett. {\bf 
 51} (1983) 1137.

\bibitem{E80} $E80$ collaboration, M.J. Alguard et al.: Phys. Rev. Lett. 
 {\bf 37} (1976) 1261; {\em ibidem} {\bf 41} (1978) 70.

\bibitem{EMC} $EMC$ collaboration, J. Ashman et al.: Nucl. Phys. {\bf B328} 
 (1989) 1.

\bibitem{HERMES} $HERMES$ collaboration, K. Ackerstaff et al.: Phys. Lett. 
 {\bf B404} (1997) 383. A. Airapetian et al.: Phys. Lett. {\bf B442} (1998) 
 484.  K. Ackerstaff et al.: Phys. Lett. {\bf B444} (1998) 531 383.

\bibitem{E155} $E155$ collaboration: P.L. Anthony et al.: Phys. Lett. 
 {\bf B493} (2000) 19.

\bibitem{Drechsel} P. Mergell, Ulf-G. Meissner and D. Drechsel: Nucl. Phys. 
 {\bf A596} (1996) 367.

\bibitem{Bianchi} N. Bianchi and E. Thomas: Phys. Lett. {\bf B450} (1999) 
 439.

\bibitem{ALLM} H. Abramowicz et al.: Phys. Lett. {\bf B269} (1991) 465.

\bibitem{Reya} B. Lampe and E. Reya: Phys. Rep. {\bf 332} (2000) 1.

\bibitem{E155_g2} $E155$ collaboration: P.L. Anthony et al.: Phys. Lett. 
 {\bf B493} (2000) 529.

\bibitem{WW} S. Wandzura and F. Wilczek: Phys. Lett. {\bf B72} (1977) 195.

\bibitem{Walker} R.L. Walker: Phys. Rev. {\bf 182} (1969) 1729.

\bibitem{Weise} J. Edelmann et al.: Nucl. Phys. {\bf A665} (2000) 125.

\bibitem{UIM} D. Drechsel et al.: Nucl. Phys. {\bf A645} (1999) 145; Phys. 
 Rev. {\bf D63} (2001) 114010.

\bibitem{Burkert} V. Burkert: Czech. J. Phys. {\bf 46} (1996) 628.

\bibitem{Thia} C. Keppel: PHD Thesis, American University, 1994.

\bibitem{BCS} H. Burkhardt and W.N. Cottingham: Ann. Phys. (N.Y.) {\bf 56} 
 (1970) 453.

\bibitem{BI} V.D. Burkert and B.L. Ioffe: Phys. Lett. {\bf B296} (1992) 223.

\bibitem{Ma} W.X. Ma et al.: Nucl. Phys. {\bf A635} (1998) 497.

\bibitem{ST} J. Soffer and O.V. Teryaev: Phys. Rev. {\bf D51} (1995) 25.

\bibitem{Wally} X. Ji and W. Melnitchouk: Phys. Rev. {\bf D56} (1977) R1.

\bibitem{Klaus} K. Helbing: in Proceedings of the IX Int'l Conference on 
 {\em Deep Inelastic Scattering}, April 27 - May 1 2001, Bologna (Italy), 
 World Scientific Publishing, in press.

\bibitem{TM} H. Georgi and H.D. Politzer: Phys. Rev. {\bf D14} (1976) 1829. 
 A. De Rujula, H. Georgi and H.D. Politzer: Ann. of Phys. {\bf 103} (1977) 
 315.

\bibitem{Carlson} C.E. Carlson and N.C. Mukhopadhyay: Phys. Rev. {\bf D47}
 (1993) R1737.

\bibitem{GRSV} M. Gl\"uck et al.: Phys. Rev. {\bf D63} (2001) 094005.

\bibitem{GS96} T. Gehrmann and W.J. Stirling: Phys. Rev. {\bf D53} (1996) 
 6100.

\bibitem{Gual} G. Altarelli: Phys. Rept. {\bf 81} (1982) 1.

\bibitem{Brodsky} S.J. Brodsky and G.P. Lepage: $SLAC$ Summer Institute on {\em Particle Physics}, July 1979, SLAC Report No. 224 (1979).

\bibitem{Amati} D. Amati et al.: Nucl. Phys. {\bf B173} (1980) 429.

\bibitem{Catani} S. Catani et al.: Nucl. Phys. {\bf B478} (1996) 273; JHEP 
 {\bf 07} (1998) 024.

\bibitem{Vogt} A. Vogt: Phys. Lett. {\bf B471} (1999) 97.

\end{thebibliography}
\end{document}